\documentclass[conference]{IEEEtran}
\IEEEoverridecommandlockouts
\usepackage{cite}
\usepackage{amsmath,amssymb,amsfonts}
\usepackage{algorithmic}
\usepackage{graphicx}
\usepackage{textcomp}
\usepackage{xcolor}
\usepackage{listings}
\usepackage[T1]{fontenc}
\usepackage{url}

\usepackage{breakurl}
\usepackage{booktabs}
\usepackage{subcaption}

\def\BibTeX{{\rm B\kern-.05em{\sc i\kern-.025em b}\kern-.08em
    T\kern-.1667em\lower.7ex\hbox{E}\kern-.125emX}}

\usepackage[breaklinks=true]{hyperref}

\begin{document}

\title{Speeding up Model Loading with fastsafetensors}

\author{\IEEEauthorblockN{Takeshi Yoshimura}
\IEEEauthorblockA{\textit{IBM Research - Tokyo} \\
Tokyo, Japan \\
tyos@jp.ibm.com}
\and
\IEEEauthorblockN{Tatsuhiro Chiba}
\IEEEauthorblockA{\textit{IBM Research - Tokyo} \\
Tokyo, Japan \\
chiba@jp.ibm.com}
\and
\IEEEauthorblockN{Manish Sethi}
\IEEEauthorblockA{\textit{IBM Research} \\
Durham, USA \\
manish.sethi1@ibm.com}
\and
\IEEEauthorblockN{Daniel Waddington}
\IEEEauthorblockA{\textit{IBM Research} \\
San Jose, USA \\
daniel.waddington@ibm.com}
\and
\IEEEauthorblockN{Swaminathan Sundararaman}
\IEEEauthorblockA{\textit{IBM Research} \\
San Jose, USA \\
swami@ibm.com}
}

\maketitle

\begin{abstract}
The rapid increases in model parameter sizes introduces new challenges in pre-trained model loading.
Currently, machine learning code often deserializes each parameter as a tensor object in host memory before copying it to device memory.
We found that this approach underutilized storage throughput and significantly slowed down loading large models with a widely-used model file formats, safetensors.
In this work, we present fastsafetensors, a Python library designed to optimize the deserialization of tensors in safetensors files.
Our approach first copies groups of on-disk parameters to device memory, where they are directly instantiated as tensor objects.
This design enables further optimization in low-level I/O and high-level tensor preprocessing, including parallelized copying, peer-to-peer DMA, and GPU offloading.
Experimental results show performance improvements of 4.8x to 7.5x in loading models such as Llama (7, 13, and 70 billion parameters), Falcon (40 billion parameters), and the Bloom (176 billion parameters).
\end{abstract}


\section{Introduction}
\label{sec:intro}

Large language models (LLMs) have enabled advanced natural language processing systems, such as chatbots~\cite{ChatGPT,Gemini} and program code generation~\cite{GraniteCode}.
The deployment of LLM systems in enterprise environments, particularly as inference servers, has seen progress in areas such as retrieval-augmented generation (RAG) and Compound AI systems~\cite{CompoundAISystem}.
Despite challenges related to managing large pre-trained models and handling high volumes of user requests,
advancements in GPU memory utilization and task scheduling~\cite{FlashAttention-NeurIPS22,FlashAttention2-ICLR24,PagedAttention-SOSP23,Orca-OSDI22,SpeculativeDecoding,DistServe,Sarathi-OSDI24} make millisecond-level token generation achievable now.

On the other hand, current LLM inference servers still face the issue of long startup times~\cite{Tensorizer,ServerlessLLM}.
The length of server startup times, which are minute-long latencies as we show in this paper, significantly disrupts user experience and server availability due to noticeably slow responses of first-token generations.
In particular, long startup times can delay or complicate software development, testing, seamless scaling, failover, and other administrative operations in practice.
This issue is further exacerbated by the rapid growth in the size of pretrained models, which now scale to billions or even trillions of parameters~\cite{Megatron-Arxiv}.
As a result, users continue to experience slow startup times, even when model files are pre-downloaded and cached on local storage.

In this work, we focus on safetensors, a widely used file format for the public distribution of pretrained models~\cite{Safetensors}.
As of March 2025, 621,008 out of 1,492,435 models (42\%) in the Hugging Face model repositories are tagged as ``Safetensors''~\cite{HuggingFace-Models}, highlighting its growing adoption in the community.
Compared to other common model distribution formats such as Pickle~\cite{Pickle} and PyTorch~\cite{Pytorch}, safetensors offers high performance and security by simplifying the serialization format.
The safety of safetensors comes from its format design, which prohibits the execution of arbitrary code upon loading serialized objects --- an issue that Pickle can potentially allow.
Safetensors enables users to efficiently test and deploy common models without the security risks.

However, we found that the current implementation of safetensors does not fully leverage the storage performance, especially on high-performance devices like NVMe SSDs.
During the startup sequence, inference servers deserialize model parameters as tensor objects provided by machine learning frameworks (e.g., PyTorch~\cite{Pytorch} and TensorFlow~\cite{TensorFlow-OSDI16}) to facilitate computation and memory management within their model architectures.
Currently, existing deserialization mechanisms instantiate tensor objects one by one in host memory before copying them to device memory.
This process results in fine-grained I/O operations and increased OS kernel load during deserialization, preventing full utilization of storage throughput, particularly in multi-GPU inference scenarios.
Additionally, the use of bounce buffering in host memory introduces redundant memory footprint and copy operations.

In training workloads, significant efforts have been made (e.g., \cite{ZeroInfinity,FDSP-Arxiv,ByteCheckpoint-Arxiv}) to optimize the serialization and deserialization of tensor objects to efficiently train large models.
Unfortunately, however, these efforts primarily focus on data movement during computation, rather than during the model initialization with safetensors.
A key challenge in this work is optimizing model loading performance while ensuring compatibility with the existing safetensors file format specification.

In this work, we design and implement \textit{fastsafetensors} as a demonstration to optimize parameter deserialization using safetensors files at local storage.
We design fastsafetensors with the existing AI software stack in mind, enabling users to easily replace their current safetensors deserialization with fastsafetensors APIs.
The main design difference from existing deserializers is that fastsafetensors decouples low-level I/O transfers from tensor object deserialization to aggregate deserialization of multiple tensors.
It directly transfers a large group of tensors from files to device memory and then instantiates tensor objects by mapping the device memory.
We take advantage of the format information in safetensors to efficiently copy larger, contiguous buffers containing multiple tensors during I/O transfers.
Additionally, we leverage DLPack~\cite{DLPack} to directly instantiate tensor objects from the raw byte buffers, eliminating the need for redundant memory copies.

Decoupling I/O transfers from object deserialization enables further optimization in low-level I/O, including parallelized copying, GPUDirect Storage (GDS)~\cite{GDS}, and NUMA awareness.
GDS enables applications to trigger DMA between storage and GPU memory with host CPU and memory bypassed while preserving file interfaces on existing Linux filesystems.
However, it requires additional software configurations including kernel modules, user libraries, filesystem type, and host system file access settings.
To accommodate environments that do not meet GDS requirements, we provide an option in fastsafetensors to fall back to the POSIX file interfaces (e.g., \verb|open| and \verb|pread|).
For instance, the fallback mode enables our I/O optimization on \verb|tmpfs|, a common DRAM-backed filesystem, which GDS does not support currently.

%However, we found that GDS introduces trade-offs between performance and configuration complexity.
%We recognize that not all environments can meet the strict requirements of GDS due to security policies and administrative limitations even though GDS provides the ``compatibility mode''.
%To address this, we provide an option in fastsafetensors to fall back to POSIX file interfaces (e.g., \verb|open| and \verb|pread|) although GDS's compatibilty mode provides similar functionality.
%This ensures compatibility with a wide range of environments while minimizing performance drawbacks.

Another feature of fastsafetensors is the offloading of tensor preprocessing tasks, such as sharding to multiple GPUs, which is common in existing inference servers.
Existing safetensors library requires user programs to perform slicing for sharding in host memory, followed by copying to GPU memory.
However, this approach increases OS kernel load due to its heavy concurrent accesses to files and page cache when multiple GPUs attempt to load them.
We simplify this file access pattern for sharding by introducing an API that leverages collective operations, such as torch.distributed's broadcast and scatter, after instantiating tensors on GPUs.

For performance evaluation, we experiment model loading of Llama~\cite{Llama-Arxiv}, Falcon~\cite{Falcon-Arxiv}, and Bloom~\cite{Bloom-Arxiv} on single and multiple GPUs.
Compared to the existing safetensors deserializer, fastsafetensors demonstrates improvements in GPU load performance ranging from 4.8x to 7.5x, while also achieving better resource utilization efficiency.
Additionally, we investigated the effectiveness of GDS in specific hardware configurations, such as scenarios where data is transferred across NUMA nodes.
To explore this, we conducted experiments with the Bloom model using 8 GPUs distributed across two NUMA nodes, with NVMe SSDs connected to only one of the nodes.
The cross-NUMA experiments revealed that GDS degraded the overall performance, highlighting the limitations of GDS in such configurations.

The contributions of this work are as follows:
\begin{itemize}
\item We present a quantitative performance analysis of current safetensors file loading, with detailed observation of resource usages under both single- and multiple-GPUs.
Our analysis indicates that the startup processes of current inference servers face bottlenecks and inefficiencies during deserialization, particularly when using high-performance storage such as NVMe SSDs.
These findings are applicable to existing serving systems~\cite{TGIS,vLLM,SGLang-Arxiv} that rely on the current safetensors deserializer.
\item This paper presents the design and implementation of the optimized model deserializer, fastsafetensors.
We aggregate tensor deserialization and offload tensor preprocessing tasks to GPUs to optimize data transfer paths during model loading.
The code is publicly available as a Python API library and distribution package at \url{https://github.com/foundation-model-stack/fastsafetensors}.
\item
The optimized architecture of fastsafetensors also enables further optimization with GDS.
While GDS has been applied in use-cases such as data preprocessing~\cite{DaliBlog} and image segmentation~\cite{OracleBlog}, our experiences, as described in this paper, offer insights on how to extend this technique to other areas.
\item
We apply fastsafetensors to existing inference server systems, the text-generation-inference server (TGIS)~\cite{TGIS} and vLLM~\cite{vLLM}, and present our experimental performance results along with a detailed analysis.
In particular, our analysis closely examines the effectiveness of GDS in practical scenarios, such as device topology awareness and co-located jobs.
We believe our findings are useful for deploying inference servers with fastsafetensors, and also offer insights into GDS characteristics.
\end{itemize}

The organization of this paper is as follows.
Section~\ref{sec:motivation} provides a detailed analysis of the current safetensors deserializer across different models and varying numbers of GPUs.
Section~\ref{sec:design} describes the design and implementation of key techniques, including aggregated tensor deserialization, GPU offloading, and GDS.
It also explains the APIs available for developers to integrate into their implementations.
Section~\ref{sec:exp} presents our experimental results and analysis.
In Section~\ref{sec:related}, we discuss different aspects of this work with related work.
Section~\ref{sec:discussion} explores practical trade-offs of GDS deployment and engineering issues as our future work.
Finally, we conclude our work in Section~\ref{sec:conclusion}.

\section{Motivation}
\label{sec:motivation}

In this section, we provide an overview of the safetensors format and present preliminary experiments with the current deserializer in safetensors 0.4.3.

\subsection{Safetensors format}
\label{sec:format}

Safetensors is a file format designed for efficiently saving and loading pre-trained models.
Unlike Pickle, which is used to save general objects, safetensors only contains tensor data and metadata such as data types, sizes, and file offsets.
The format is simplified to allow zero-copy reading to enable tensor instantiation in host memory using mmap.

\begin{figure}[t]
\centering
\includegraphics[width=0.6\linewidth]{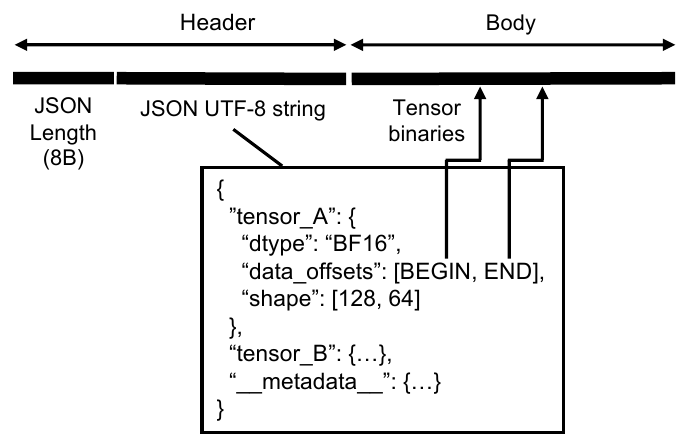}
\caption{Overview of the safetensors format: The file is divided into a header and a body, with its layout defined by a JSON string.}
\label{fig:safetensors_format}
\end{figure}

Figure~\ref{fig:safetensors_format} illustrates the layout of the safetensors format.
A safetensors file is divided into two sections: a header and a body for serialized tensors.
The header begins with 8 bytes to store its size, followed by a JSON-formatted layout definition for each tensor.
Each tensor is associated with a key so that users can easily identify and deserialize specific tensors in the body section.
The body section contains serialized tensors at offsets specified in the layout definition.

%\begin{figure}[t]
%\begin{lstlisting}
%with safe_open("a.safetensors") as f:
%    for key in f.get_keys():
%        t = f.get_tensor(key)
%\end{lstlisting}
%\caption{Example of reading a safetensors file.}
%\label{fig:safetensors_safeopen}
%\end{figure}

The safetensors library enables users to easily deserialize tensors in a file.
It parses the header and prepares mapping offset ranges corresponding to a requested tensor key in host memory as needed.
Users can call \verb|get_tensor()| with a key to deserialize a tensor object in host memory and perform preprocessing tasks such as type/layout conversions, slicing, concatenation, tranposistion, and copy to device memory.
The library also provides a special function, \verb|get_slice()|, to deserialize only part of an on-disk tensor.

% TODO 1-(b): Figure 2a shows the performance of the current safetensor deserializer using the same strategy as TGIS. Could you briefly describe this strategy to help the reader better understand how deserialization is performed in TGIS?
TGIS and vLLM leverages the function to partition a tensor for tensor parallelism to efficiently distribute LLMs across multiple GPUs.
Each transformer layer's parameters are partitioned across devices: input projection weights such as those in the self-attention query, key, and value (QKV) layers and the feed-forward multi-layer perceptron (MLP) are split columnar-wise, while output projection weights are split row-wise.
Within a node, inference processes can leverage the OS kernel to unify file reads at the virtual filesystem layer, allowing efficient deserialization of partitioned tensors.

\subsection{Performance Analysis}
\label{sec:motivation:analysis}

\begin{table}[t]
\caption{Basic statistics of used models.}
\label{table:models}
%\vskip 0.15in
\begin{center}
\begin{scriptsize}
\begin{sc}
\begin{tabular}{lccc}
\toprule
Model-parametersize & \# of files & Size (GB) \\
\midrule
Llama-7B~\cite{Llama-Arxiv} & 2 & 13 \\
Llama-13B~\cite{Llama-Arxiv} & 3 & 24 \\
Falcon-40B~\cite{Falcon-Arxiv} & 9 & 78 \\
Llama-70B~\cite{Llama-Arxiv} & 15 & 128 \\
Bloom-176B~\cite{Bloom-Arxiv} & 72 & 328 \\
\bottomrule
\end{tabular}
\end{sc}
\end{scriptsize}
\end{center}
\vskip -0.1in
\end{table}

Our preliminary experiment uses models including Llama-7B, Llama-13B, Llama-70B, Falcon, and Bloom (Table~\ref{table:models}).
The test environment consists of Red Hat Core OS 4.14 (Linux 4.18) with 80 Icelake CPUs and 1.2 TB of DRAM.
All the models are stored as safetensors files on four 3.2-TB NVMe SSDs formatted as XFS.
We load Llama-7B and Llama-13B using a single A100 GPU, Falcon-40B using two GPUs, Llama-70B using four GPUs, and Bloom using eight GPUs.
For multi-GPU loading, we start and synchronize the loading processes using PyTorch's distributed mode with the NCCL backend, with the safetensors files evenly-distributed across the four NVMe SSDs.
We partition models into safetensors files of 5 GB to 10 GB, assigning an equal number to each rank by creating symbolic links in model cache directories.

\begin{figure}[t]
\centering
\begin{subfigure}{0.34\linewidth}
\centering
\includegraphics[width=\linewidth]{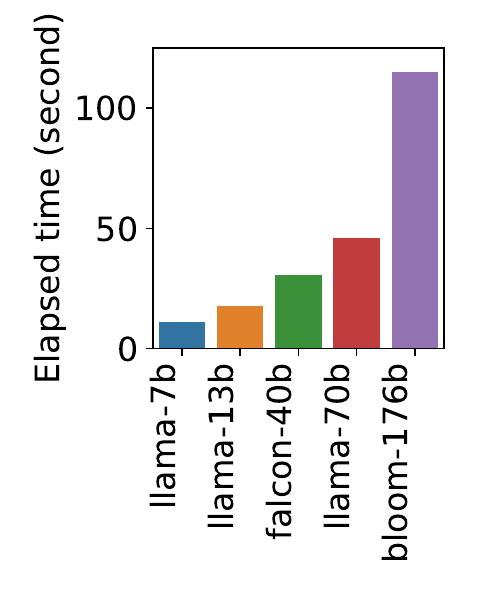}
\caption{Elapsed time.}
\label{fig:mmap_elapsed_time}
\end{subfigure}
\hfill
\begin{subfigure}{0.28\linewidth}
\centering
\includegraphics[width=\linewidth]{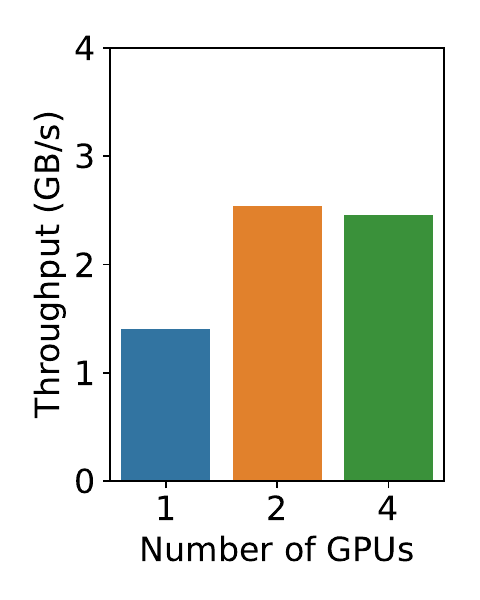}
\caption{Strong scaling of Falcon.}
\label{fig:mmap_strong_scaling}
\end{subfigure}
\hfill
\begin{subfigure}{0.28\linewidth}
\centering
\includegraphics[width=\linewidth]{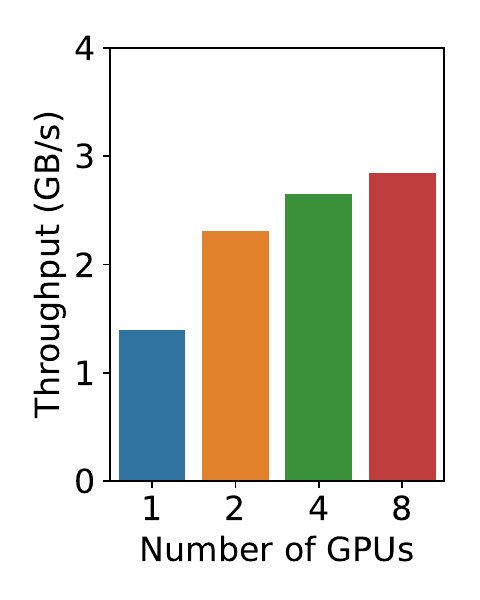}
\caption{Weak scaling of Bloom.}
\label{fig:mmap_weak_scaling}
\end{subfigure}
\caption{Performance of current safetensors deserializer.}
\label{fig:mmap_perf}
\end{figure}

Figure~\ref{fig:mmap_elapsed_time} shows the performance of the current safetensors deserializer when loading different models from four NVMe SSDs.
The experiments deserialize and partition tensors using the same strategy as TGIS as described in Section~\ref{sec:format}.
The figure shows that model loading required more than several seconds to even minutes, depending on model size.
Note that the maximum throughput of the four NVMe SSDs was 28 GB/s in total (7 GB/s per SSD) under our stress tests.
However, the estimated throughput during model loading was significantly lower than the storage's maximum throughput.

We also experiment with strong scaling of Falcon by varying concurrency on the same file size (Figure~\ref{fig:mmap_strong_scaling}) and weak scaling of Bloom by adjusting file sizes to be proportional to concurrency (Figure~\ref{fig:mmap_weak_scaling}).
These figures indicate that the current deserializer does not scale well, resulting in loading times of more than a minute for Bloom on eight GPUs.

\begin{figure}[t]
\centering
\begin{subfigure}{0.48\linewidth}
\centering
\includegraphics[width=\linewidth]{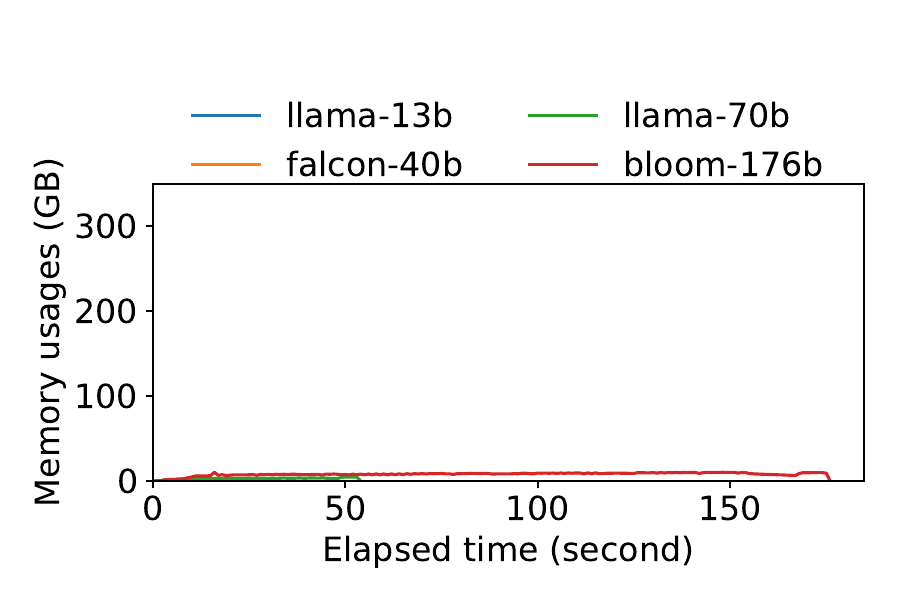}
\caption{User memory.}
\label{fig:mmap_used}
\end{subfigure}
\hfill
\begin{subfigure}{0.48\linewidth}
\centering
\includegraphics[width=\linewidth]{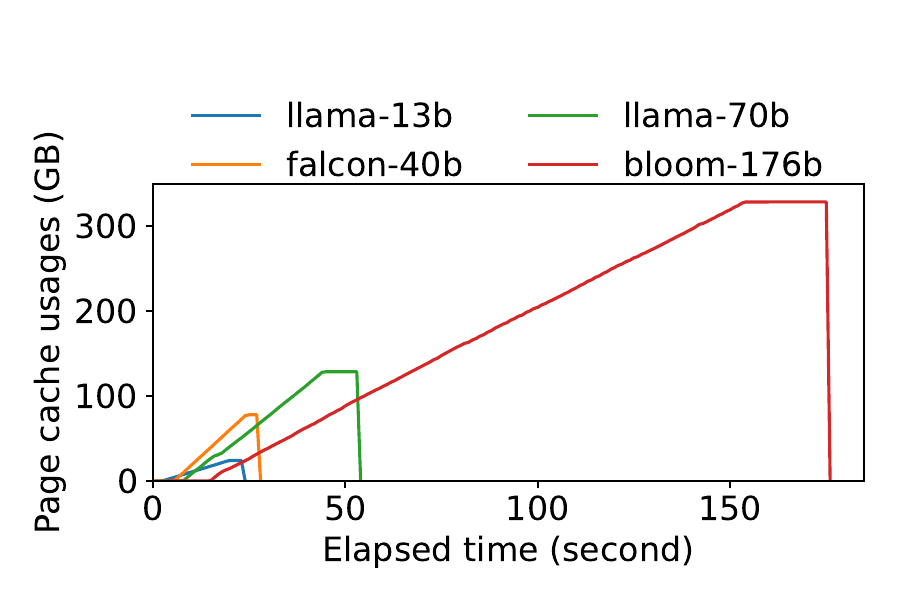}
\caption{Page cache.}
\label{fig:mmap_cach}
\end{subfigure}
\vfill
\begin{subfigure}{0.48\linewidth}
\centering
\includegraphics[width=\linewidth]{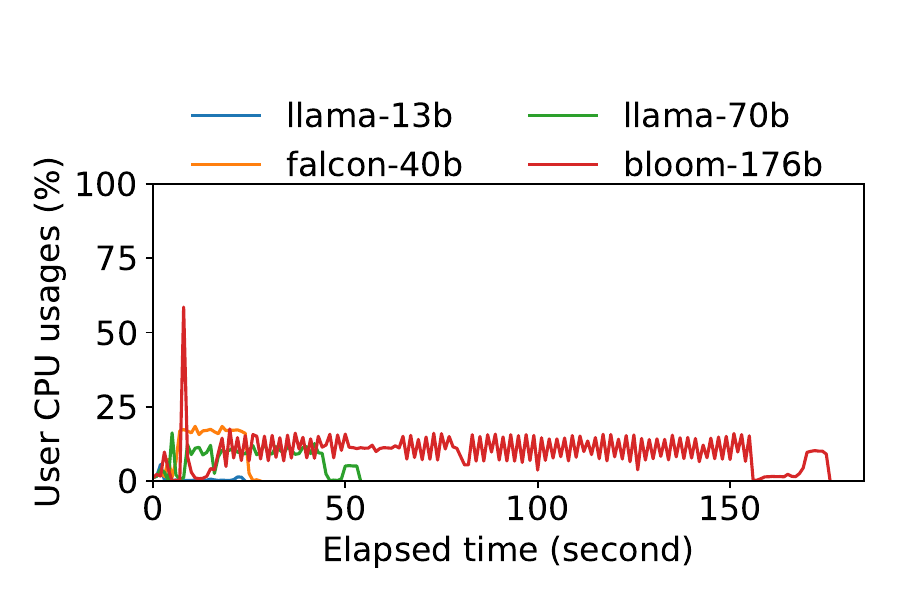}
\caption{User CPU.}
\label{fig:mmap_usr}
\end{subfigure}
\hfill
\begin{subfigure}{0.48\linewidth}
\centering
\includegraphics[width=\linewidth]{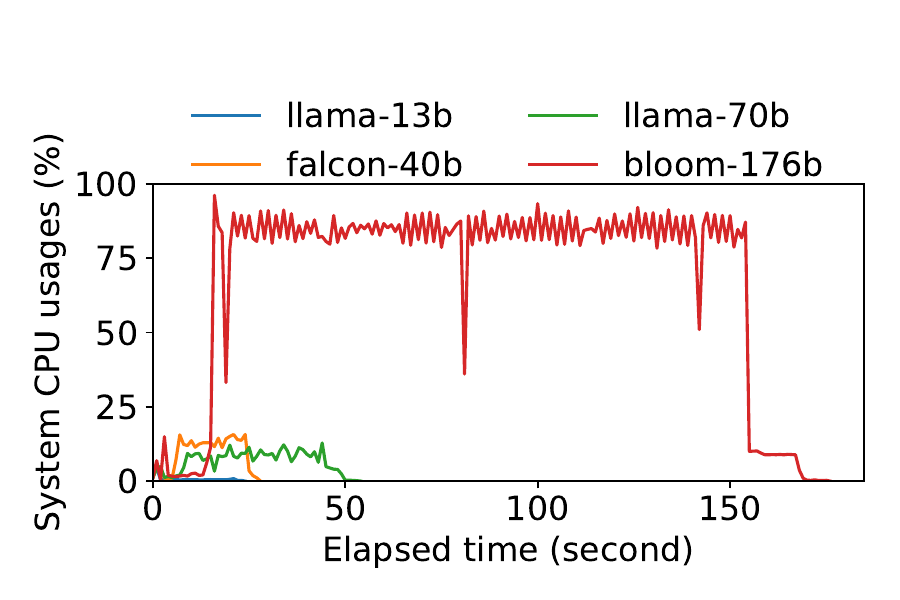}
\caption{Kernel CPU.}
\label{fig:mmap_sys}
\end{subfigure}
\vfill
\begin{subfigure}{0.48\linewidth}
\centering
\includegraphics[width=\linewidth]{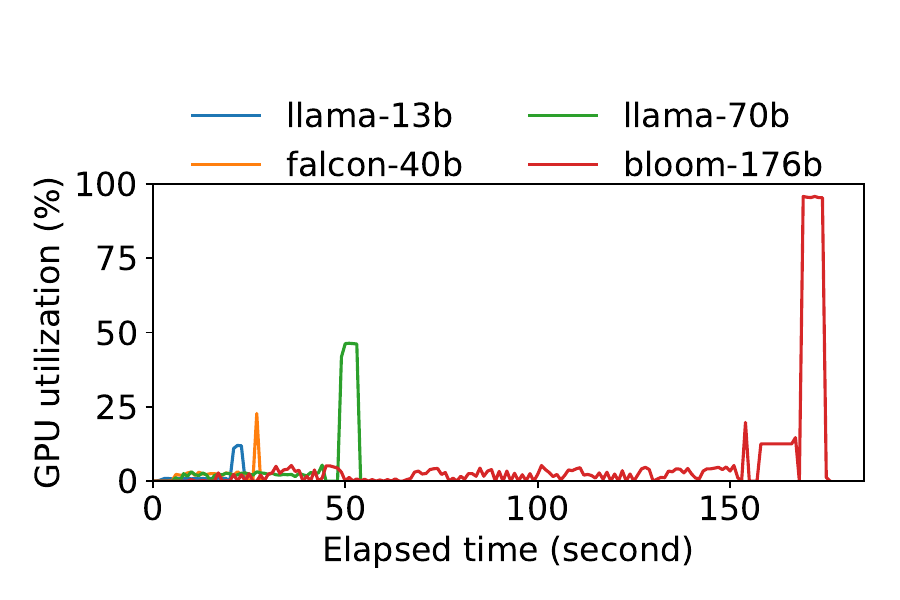}
\caption{GPU utilization.}
\label{fig:mmap_util}
\end{subfigure}
\hfill
\begin{subfigure}{0.48\linewidth}
\centering
\includegraphics[width=\linewidth]{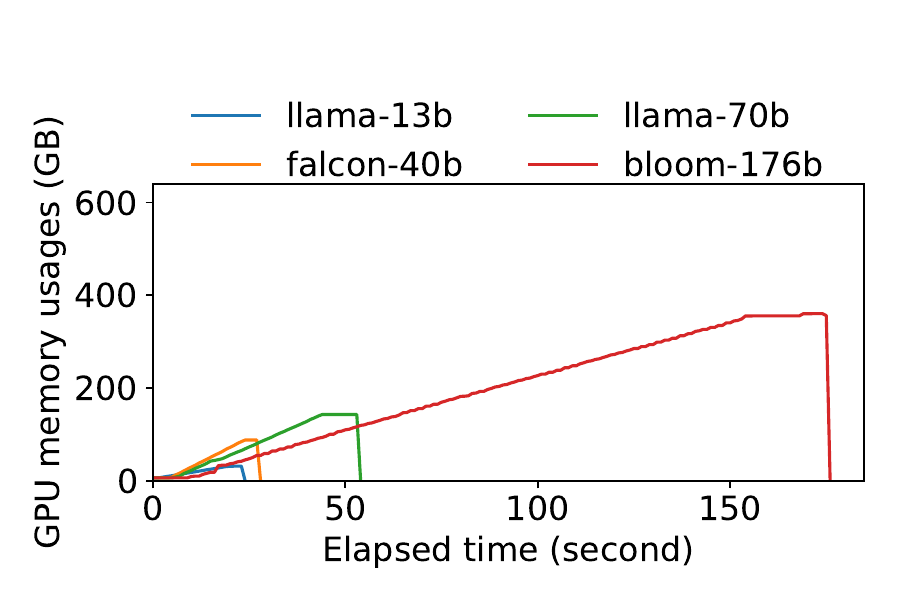}
\caption{GPU memory.}
\label{fig:mmap_gpumem}
\end{subfigure}
\vfill
\begin{subfigure}{0.48\linewidth}
\centering
\includegraphics[width=\linewidth]{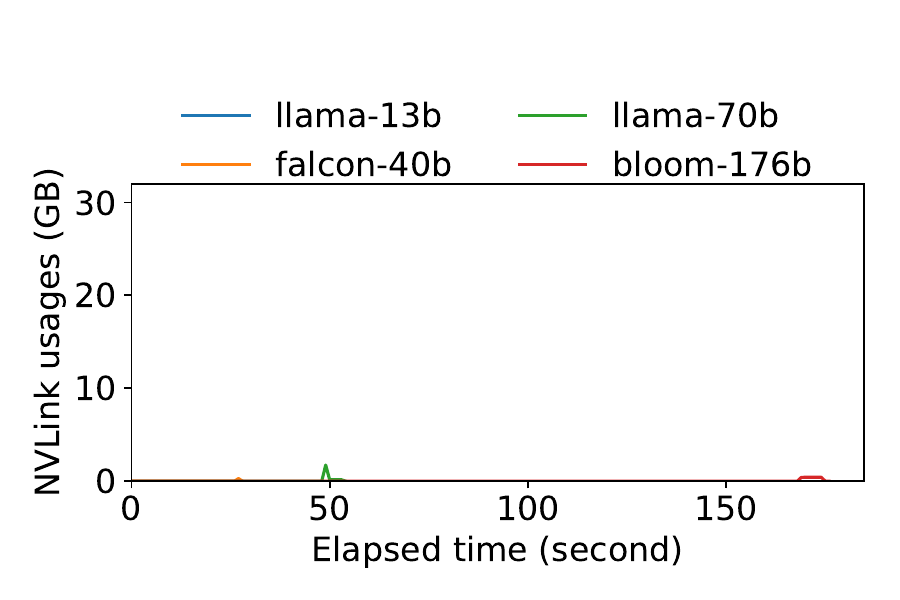}
\caption{NVLink.}
\label{fig:mmap_nvl}
\end{subfigure}
\hfill
\begin{subfigure}{0.48\linewidth}
\centering
\includegraphics[width=\linewidth]{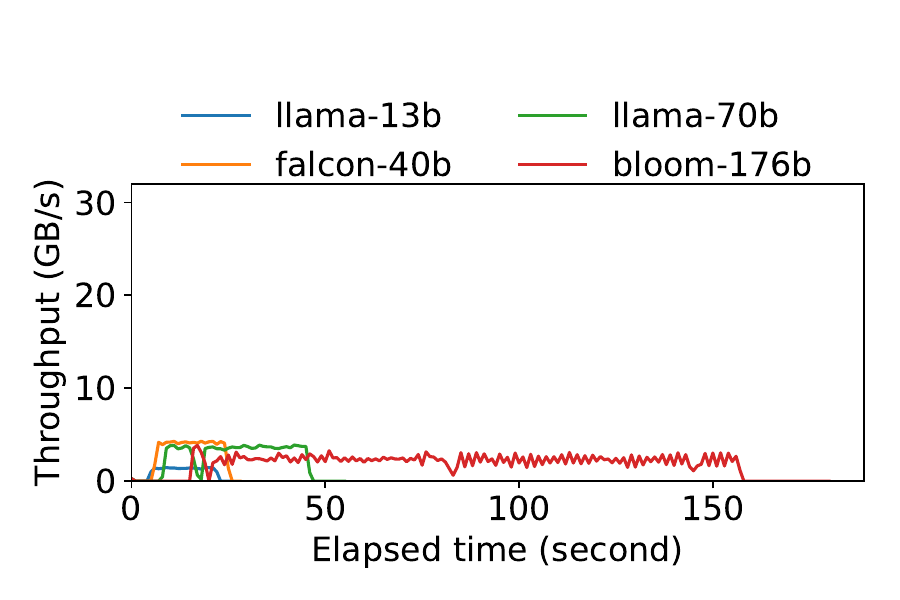}
\caption{NVMe SSD read.}
\label{fig:mmap_disk}
\end{subfigure}
\caption{Resource utilization of TGIS.}
\label{fig:current_srcutil}
\end{figure}

Next, we run TGIS and analyze its resource utilization with different models until the first generation sequence (128 tokens) completes.
In our experiments, which used 263 input tokens in 1.2 KB texts, model loading accounted for an average of 92\% of the startup latency.
Figure~\ref{fig:current_srcutil} illustrates the resource usages during experiments with different models.

Figure~\ref{fig:mmap_used} and \ref{fig:mmap_cach} illustrate the large footprint of the page cache during model loading, while showing relatively less usagses of host memory.
The sizes of page cache usage were consistent with the sizes of model files (for example, 328 GB for Bloom).
The heavy usages of the page cache led to a high kernel CPU load, as described in Figure~\ref{fig:mmap_sys}.
Bloom exhibited notably high utilization for concurrent accesses to the page cache because the GPU driver locks the memory pages for DMA.
Figure~\ref{fig:mmap_usr} shows relatively low but still significant user CPU utilization, especially for loading Falcon-40B, Llama-70B, and Bloom, which involve partitioning tensors for sharding.
Machine learning code that uses the safetensors library needs to partition tensors within the code, increasing memory accesses and copies, along with host CPU usages.

In contrast, GPU resources are not fully utilized during model loading as shown in Figure~\ref{fig:mmap_util} and \ref{fig:mmap_nvl}.
We observed GPU and NVLink utilization only after starting inference for all the models.
The GPU memory usages incrementally increased at the same pace as page cahce usages, as illustrated in Figure~\ref{fig:mmap_gpumem}.

Figure~\ref{fig:mmap_disk} shows that the four NVMe SSDs, which have a maximum total throughput of 28 GB/s, were utilized at a maximum of 5 GB/s for all the models.
This throughput difference is attributed to the fact that multiple processes read files in parallel.

\subsection{Design observation}

These preliminary results reveal several design issues that require improvement in the current safetensors file loading process.
Upon a careful review of its code, we identified inefficiencies in the designs for tensor deserialization and preprocessing at the host CPU and memory.
The details of these issues are as follows.

\begin{figure}[t]
\centering
\includegraphics[width=0.7\linewidth]{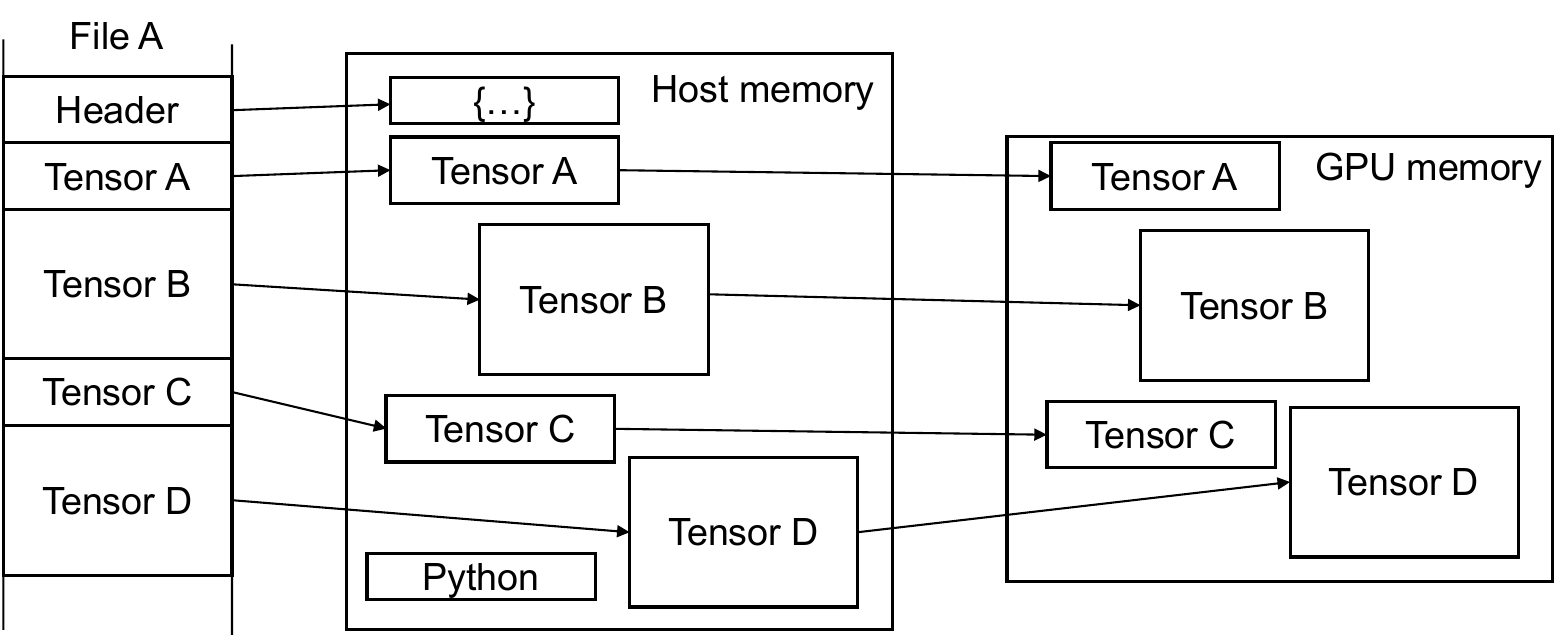}
\caption{Tensor copy flow.}
\label{fig: current_copy}
\end{figure}

\textbf{Issue 1: Instantiating tensors one by one in host memory (Figure~\ref{fig: current_copy}): }
In the current safetensors approach, tensors within the file are mapped to host memory, instantiated, and then copied to GPU memory.
Existing inference servers are often implemented in Python, where tensor instantiation is processed sequentially.
Furthermore, due to the on-demand loading of files via mmap, file prefetching relies on generic heuristics in Linux, which makes it challenging to achieve optimal performance with high-performance storage such as NVMe.
%TODO 1-(c): Issue 1 appears to be more related to a Python limitation—the infamous Global Interpreter Lock (GIL), which you address through a CPython binding. CPython 3.13 introduces experimental support for running without the GIL. Could you briefly comment on the extent to which Issue 1 could potentially be mitigated in the latest version of Python?
%A fundamental issue here is that the current safetensors approach does not decouple file reads from object instantiation but need doing them at once.
Python GIL removal~\cite{Python-GIL} would enable us to parallelize tensor instantiation, but we expect that it still cannot fully utilize the storage bandwidth due to increased coordination overhead within the Python runtime for object management.

\begin{figure}[t]
\centering
\includegraphics[width=0.7\linewidth]{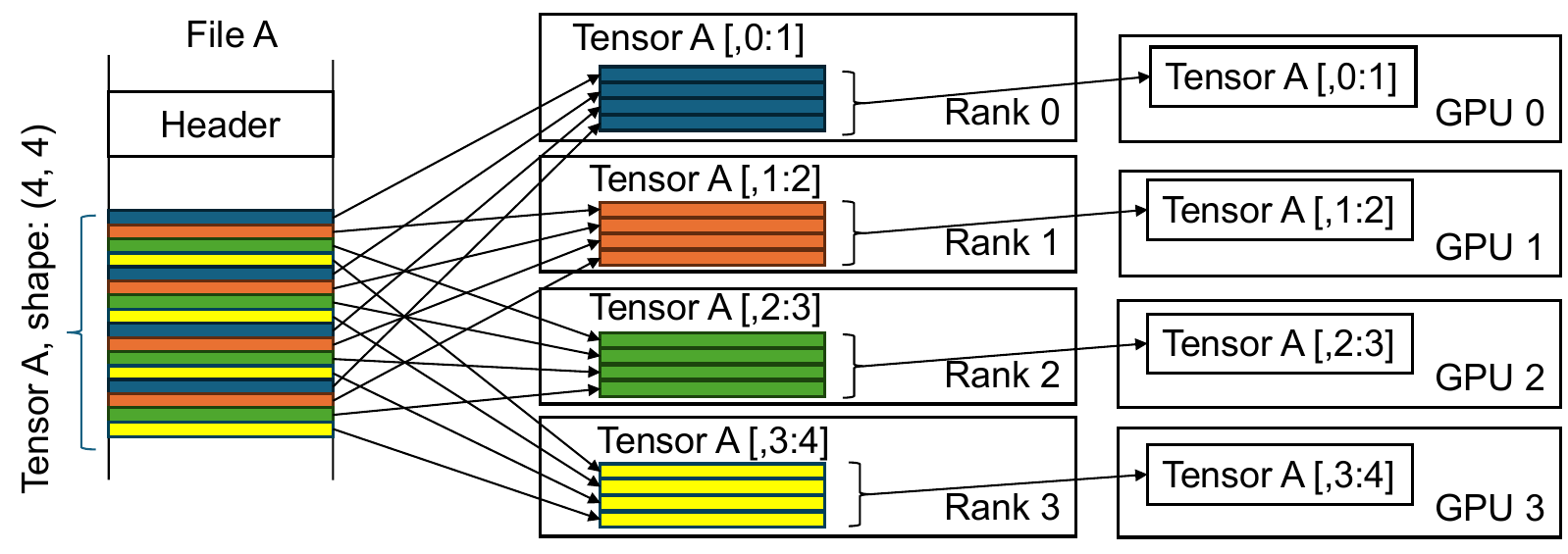}
\caption{Tensor sharding.}
\label{fig: current_shard}
\end{figure}

\textbf{Issue 2: Preprocessing on the host CPU (Figure~\ref{fig: current_shard}): }
when performing tensor parallelism across multiple GPUs, the current safetensors library requires users to obtain split tensors in host memory and then copy them to GPU memory.
However, this approach leads to excessive copying of small tensors, resulting in significant overhead for DMA operations and slowing down the loading process.
Additionally, this procedure unnecessarily increases OS kernel load related to page cache, further slowing down the entire workload, as illustrated in the resource utilization of Bloom.

\textbf{Issue 3: Inefficient memory allocation: }
When using pre-trained models that are typically distributed, nearly all of the weights in the file need to be loaded onto the GPU.
However, the current safetensors approach maps the entire content of the file into the user process's virtual memory, consuming host memory equivalent to the size of the weights.
Despite this, in actual inference servers, the weights are further copied to the GPU, meaning that the host memory consumption is only required for the loading process.

\section{Fastsafetensors}
\label{sec:design}

In this section, we present our design and implementation of fastsafetensors to optimize model loading for efficient inference server startups.
Fastsafetensors enables us to highly utilize storage throughput by transferring a large group of on-disk tensors to GPU memory with aggregated tensor deserialization (Section~\ref{sec:design:tensor}).
It also performs GPU offloading for preprocessing tasks on tensors, such as sharding and type conversion (Section~\ref{sec:design:preprocess}).
We describe its APIs, which differ from the file loading APIs of the original safetensors library when using multiple GPUs (Section~\ref{sec:design:api}).
Additionally, we outline the end-to-end deserialization protocol of fastsafetensors.
% TODO 2-1: In the very last line in the discussion section, the authors mention that while the current implementation is for Nvidia GPUs and CUDA, the problem that fastsafetensors addresses applies to other accelerators as well. This deserves mention earlier in the paper, possibly in Section 3, where it will be useful to highlight the specific capabilities/APIs in CUDA that are utilized, so that the reader has an idea of what alternate accelerators and their software stacks need to support to extend fastsafetensors to those.
We prototype fastsafetensors with CUDA memory APIs, cuFile APIs, and collective operations such as broadcast and scatter in torch.distributed.
%However, our design assumes storage APIs for reading on-disk objects and copying them between storage and DRAM or device memory.

\subsection{Aggregated Tensor Deserialization}
\label{sec:design:tensor}

\begin{figure}[t]
\centering
\includegraphics[width=0.7\linewidth]{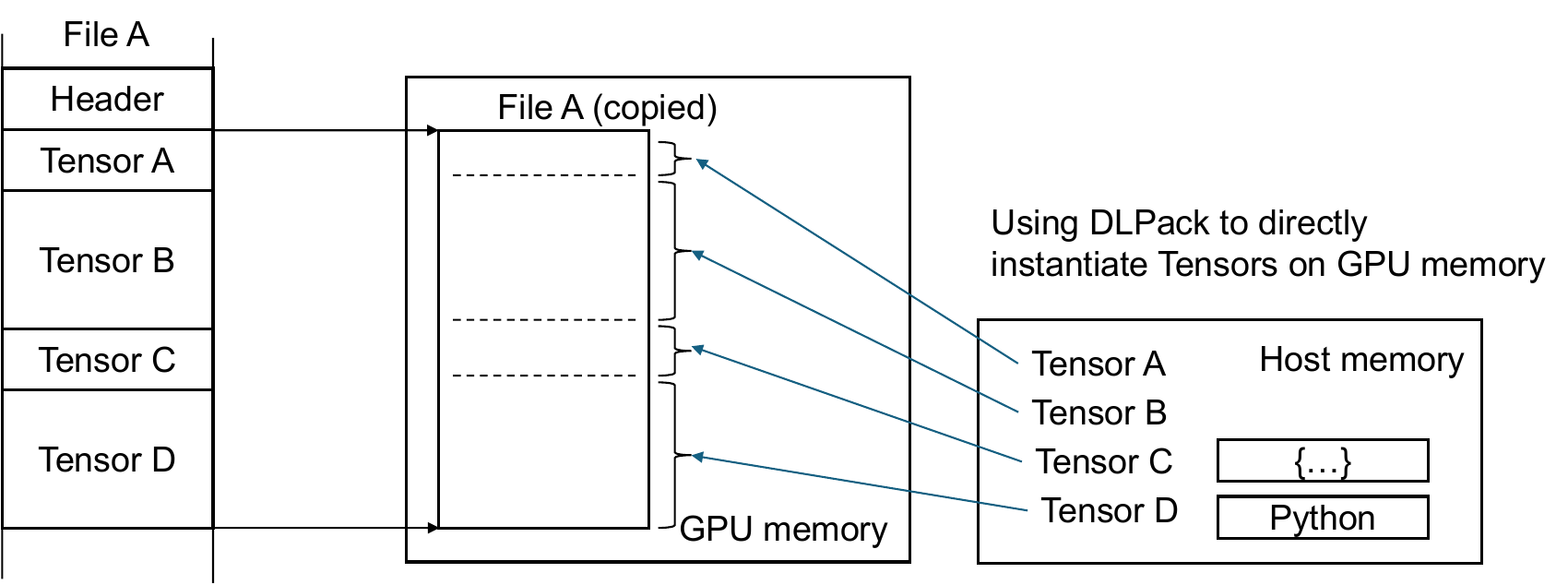}
\caption{Batching of file I/O and tensor instantiation with DLPack.}
\label{fig: approach_copy}
\end{figure}

Fastsafetensors modifies the execution flow from file I/O to tensor instantiation (see Figure~\ref{fig: approach_copy}).
We leverage the safetensors format, which serializes all the tensors in contiguous region of a file, with known offsets for each type, shape, and data in advance.
Unlike the orignal safetensors library, we first transfer a group of tensor data from the file to GPU memory and then uses DLPack~\cite{DLPack} to instantiate them as tensor objects.
%DLPack enables zero-copy format exchanges across multiple frameworks, allowing us to directly instantiate tensors from device memory.
% TODO 1-(d): It is unclear how DLPack enables the instantiation of raw tensor bytes as a tensor object directly on the GPU. More detail is expected here, as this is the fundamental contribution of the paper.
With the DLPack protocol, we wrap a cotiguous buffer as a tensor object of a given framework, specifying a raw pointer to the buffer, the target device, data type, and the strides (i.e., the number of bytes to skip for each dimension).
We compute the strides based on the given data type and shape, assuming the memory layout used by PyTorch tensors.

The primary advantage of delaying tensor instantiation is that it allows us to decouple the low-level details of data transfers from tensor object boundaries.
This decoupling enables us to implement batch processing and parallelization of file I/O with NUMA awareness.
Fastsafetensors identifies the NUMA nodes associated with NVMe SSDs and GPUs, allocating I/O threads and memory at closely as possible to the same node.
We calculate the total size of the files and partition them into transfer blocks to efficiently utilize the configured number of I/O threads.

The modified execution flow is a key enabler of GDS for deserializing various sizes of tensors in a file.
Fastsafetensors is a Python library that internally includes C-Python binding code, allowing Python code to easily invoke CUDA and cuFile APIs, which GDS offers for performing file copying to and from GPU memory.
GDS enables us to bypass the host CPU and memory, thereby fully utilizing storage throughput.
However, existing work~\cite{SPIN-ATC17} shows that GDS performance can degrade when the transfer block size is small.
Therefore, we configure the number of I/O threads to match the number of files to ensure that transfer sizes are large enough, unless this would exceed 80\% of the physical CPUs in a NUMA node.

Despite its effectiveness, we regard GDS as an optional feature because some platforms and filesystems do not meet the software and/or hardware requirements for GDS.
Thus, we also provide a fallback method that utilizes pread and cudaMemcpy with a small, DMA-enabled bounce buffer in host memory to facilitate DMA transfers to GPU memory.
Note that we did not use the compatibility mode of GDS for the fallback due to practical issues such as version requirements for CUDA and GPU models.
However, we believe our fallback method exhibits similar behavior to the mode provided by NVIDIA.
This design enables us to deploy inference servers in a broad range of environments, including caching files in DRAM using tmpfs, as we analyze in Section~\ref{sec:exp:add}.

\subsection{GPU offloading}
\label{sec:design:preprocess}

\begin{figure}[t]
\centering
\includegraphics[width=0.8\linewidth]{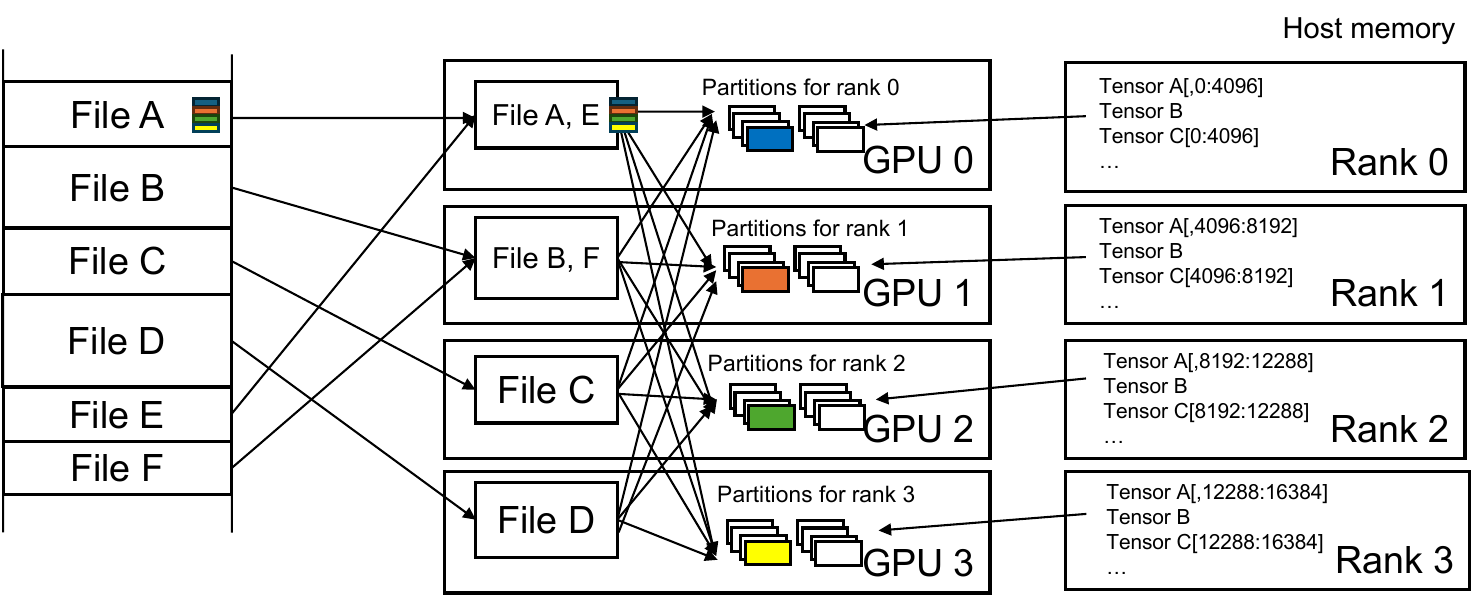}
\caption{Offloading of sharding to GPUs.}
\label{fig: approach_shard}
\end{figure}

Aggregated tensor deserialization introduces another challenge when implementing inference for multiple GPUs.
A straightforward approach could be to select the file range needed for each rank to transfer to GPU memory.
However, this naive approach would result in many small, scatterred I/O operations, leading to significant performance degradation.
Inference server implementations can also change tensor layouts and type conversions, which increase GPU memory overhead.

To overcome these issues, fastsafetensors offloads tensor preprocessing to GPUs, leveraging high-bandwidth memory and low-latency interconnects (particularly NVLink).
We provide APIs for typical sharding by partitioning tensors at a dimension to be the same shape.
In the APIs, fastsafetensors first reads each file onto different GPUs in a round-robin manner and then scatters the partitioned tensors to the GPUs used during inference (Figure~\ref{fig: approach_shard}).
Broadcasting is also available for tensors required by all ranks.
We call the phase of broadcasts and scatters among GPUs \textit{``shuffling''}.

Fastsafetensors internally tracks the shuffled tensors and return GPU memory to the Pytorch memory pool if possible.
Inference servers with tensor parallelism often iteratively initialize tensors for each layer.
Additionally, shuffling allocates new tensors to copy values from remote GPU memory, and the original tensors that are deserialized from storage become unused after shuffling.
hus, fastsafetensors provides an option to automatically release the GPU memory allocated for deserialization after shuffling, so that application code can reuse this memory to iteratively construct tensors for each layer.

The safetensors format does not define data alignment and can cause alignment errors due to constraints of both GDS and CUDA kernels.
For example, when the header of a safetensors file is odd-sized, GDS still has to transfer tensors at the odd offset within GPU memory due to its 512-byte alignment for direct data transfer.
In this case, CUDA kernels encounter alignment errors when accessing the tensors that GDS copied.
Thus, fastsafetensors redundantly copies tensors to correct the address alignment after transferring data from storage to GPU memory with GDS.
It temporarily allocates a bounce buffer in GPU memory and iteratively copies tensors back and forth to fix their offsets within the GPU memory that GDS used.

We observed that some publicly available models, including Bloom and Llama, contain odd-sized headers in their safetensors files, which require the misalignment fixes.
However, our experiments did not show significant overhead from the additional copies due to memory access locality and high-performance copying within GPU memory.
Inference servers sometimes need to convert data types (e.g., BFloat16 to half precision), and we also handle data conversions in the same process as the alignment fixes, i.e., iteratively copying via a bounce buffer.

\subsection{API and Usage Examples}
\label{sec:design:api}

\begin{figure}[t]
\begin{lstlisting}
import torch
from fastsafetensors import SafeTensorsFileLoader
from fastsafetensors import SingleGroup

device = torch.device("cuda:0")
loader = SafeTensorsFileLoader(SingleGroup(), device)
loader.add_filenames({0: ["a.safetensors"]})
fb = loader.copy_files_to_device()
tensor_a0 = fb.get_tensor("a0")
print(f"a0: {tensor_a0}")
fb.close()
loader.close()
\end{lstlisting}
\caption{Example of single GPU application. It reads a.safetensors to show a tensor a0.}
\label{fig:mprog_single}
\end{figure}

\begin{figure}[t]
\begin{lstlisting}
import torch
import torch.distributed as dist
from fastsafetensors import SafeTensorsFileLoader

dist.init_process_group(backend="gloo")
pg = dist.group.WORLD
device = torch.device("cuda:0")
loader = SafeTensorsFileLoader(pg, device)
filemap = {
    0: ["a.safetensors"], 1:["b.safetensors"],
}
loader.add_filenames(filemap)
fb = loader.copy_files_to_device()
tensor_a0 = fb.get_tensor("a0")
tensor_b0_sharded = fb.get_sharded("b0", dim=1)
print(f"RANK {pg.rank()}: {tensor_a0}")
print(f"RANK {pg.rank()}: {tensor_b0_sharded}")
fb.close()
loader.close()
\end{lstlisting}
\caption{Example of multi-GPU application. It reads files to show a tensor key a0 and shard a tensor key b0 per rank.}
\label{fig:mprog_multi}
\end{figure}

Figures~\ref{fig:mprog_single} and~\ref{fig:mprog_multi} provide code examples of file loading with fastsafetensors for one GPU and multiple GPUs, respectively.
In these examples, tensor keys a0 and b0 are included in two separate safetensors files.
Note that we describe the current low-level APIs as of this writing, but we plan to provide higher-level APIs to hide low-level details, including memory management and file mapping.

The \verb|SafeTensorsFileLoader| is the main entry point of fastsafetensors.
Applications with a single GPU can use \verb|SingleGroup()| to initialize the file loader.
For multiple GPUs, they can initialize the file loader with a \verb|ProcessGroup| from torch.distributed.
The loader supports both CPU and CUDA devices, and users can optionally enable GDS.

Then, users need to map the target files to ranks with \verb|add_filenames()|.
For a single GPU, applications only need to pass a dictionary that maps key 0 to a list of all files.
For multiple GPUs, applications need to create a list that is as evenly distributed as possible.
Currently, fastsafetensors leaves the task of list creation to the developer.
% TODO 1-(e): The Fastsafetensor API is still not user-friendly, as it requires users to manually distribute the safetensors files across the GPUs. This may not be trivial for end users. Could you clarify what would happen if the files are not properly distributed?
The list with skewed data placement may degrade performance, and thus, our future work is to provide helper APIs to prepare and relocate data at multiple storage.

Next, calling the \verb|copy_file_to_device()| method triggers file transfers to GPU memory fragments, where tensors are mapped.
It returns \verb|FilesBufferOnDevice|, which holds a group of tensor objects and exposes APIs to retrieve and shard tensors.
At this point, tensors are only available on the ranks specified in \verb|add_filenames()|.

Once the files are loaded, applications can use the \verb|get_tensor()| method to retrieve tensors.
Additionally, applications can use the \verb|get_sharded()| method to retrieve sharded tensors.
In the case of multiple GPUs, these methods internally perform collective communication using torch.distributed.

In other words, fastsafetensors assumes that the same code is executed in the same order across all ranks.
This ensures that collective communication operations are called on all ranks, enabling tensor exchange between them.
This is a common assumption for inference server implementations of tensor parallelism~\cite{TGIS,vLLM}.

As described in Section~\ref{sec:design:preprocess}, the loader tracks the refrence count of keys and releases GPU memory used by \verb|copy_file_to_device()| by default.
Users can disable this behavior at loader initialization, typically when running single-GPU inferences, to maintain tensor objects in the GPU memory used during data transfer.
To ensure the freeing of the GPU memory allocated for the data transfer, application code needs to explicitly call the \verb|.close()| method.
Therefore, it is the user's responsibility to ensure that all tensors are properly released before calling \verb|.close()|.
The \verb|.close()| method safely releases the underlying GPU memory.

\section{Evaluation}
\label{sec:exp}

In this section, we characterize the performance and resource efficiency of fastsafetensors.
We quantify the effectiveness of key optimizations, including optimized data transfers with aggregated tensor deserialization, shuffling with GPU offloads, and GDS.
Experiments conducted on a single GPU provide insights into the effectiveness that is largely derived from the benefits of aggregated tensor deserialization.
Resource observation from multi-GPU experiments allow us to better understand the benefits of shuffling.
The characteristics of GDS have already been explored in existing work~\cite{SPIN-ATC17,P2PDMA-APSys20}, but our analysis specific to model loading offers implications for its practical use-cases.
Specifically, we explore the effectiveness of device topology, co-located workloads, and advantages over DMA via bounce buffers in host memory.

\subsection{Data}

Our analysis focuses on three different models: Llama, Falcon, and Bloom, each with a different number of parameters.
These models are available as multiple safetensors files (Table~\ref{table:models}) that contain tensors from a pre-training checkpoint (they are serialized as the order of layers).
TGIS can automatically download files for pre-trained models from model repositories at HuggingFace; however, we evenly distributed the required safetensors files to four NVMe SSDs or tmpfs in advance of each experiment.
We maximize PCI Express bus utilization by allowing each rank to read a balanced number of files from the four NVMe SSDs.
Additionally, we evaluate the influences of NUMA topologies and strong/weak scaling by changing which rank loads files from which NVMe SSD.

%TODO 1-(a): First of all, it is unclear whether replicating the model across four NVMes is realistic in practice. To be more resource-efficient, the safetensor files should be evenly distributed across the NVMe SSDs, and the effect of no replication should be investigated. Could you comment on this?
Our experimental setup did not have storage capacity issues, and thus, we simply replicated all the safetensors files to four NVMe SSDs and used symbolic linking so that each rank can read the equal number of files.
However, users are not required to replicate all files to all storage devices in their real environments.
Instead, they can distribute files across different storage units corresponding to the target ranks and make their symbolic links on the cache directory.
This setup allows multi-GPU loading to be achieved easily, without relying on shared storage or additional synchronization mechanisms.

For TGIS experiemnts, we use a single-shot prompt with 1.2 KB of text to test whether fastsafetensors does not consume unnecessarily large GPU memory, which could cause out-of-memory errors during main inference computation.
The prompt is a request for a document summary of a random article about the history and definition of artificial intelligence.

\subsection{Workload}

Our experiments run on our modified TGIS and a standalone loader.
TGIS experiemnts provide more concrete insights into how fastsafetensors influence overall inference workloads,
while the standalone loader leads to a detailed understanding of the key optimizations in fastsafetensors.

Our modified TGIS invokes either the fastsafetensors APIs or the default safetensors library at its weight loader class.
The weight loader class retrieves tensors associated with specified keys from safetensors files.
It also supports sharding of retrieved tensors for different number of GPUs.
Implementations of each model architecture use the weight loader to refill parameters and biases to match the state at the time of a training checkpoint.

The standalone loader follows the same behavior as TGIS model initialization, including NCCL-based torch distributed processing and sharding policy, i.e., partitioned dimension.
As described in the previous section, fastsafetensors first copies files to GPU memory and then relocates partitioned tensors with shuffling.
In contrast, the default safetensors library utilizes \verb|mmap| to lazily load tensors with partitioned shapes into host memory from files.
It then copies the tensors to GPU memory for each rank.

Each rank reads files with a maximum of 16 threads.
Fastsafetensors with GDS disabled uses 160-MB bounce buffer in host memory for each thread.

\subsection{Experimental setup}
\label{sec:exp:overall}

The experimental setup is the same as that of the preliminary experiments described in Section \ref{sec:motivation}.
We use a single virtual server with 1.2 TB of host memory and two NUMA nodes, each consisting of 40 Intel Icelake CPUs with a total of 5 MB L1, 160 MB L2, and 32 MB L3 caches.
Each NUMA node is connected to a PCI Express Gen4 bus with four NVIDIA A100 GPUs, each having 80 GB of GPU memory.
Every GPU is interconnected via NVLink 2.0.
Additionally, each NUMA node is connected to four Samsung 3.2 TB NVMe SSDs, and thus, one set of four GPUs are closer to storage than the other due to the NUMA toplogy.
Our experiments primarily run in the NUMA node connected to the NVMe SSDs, but we also utilize this non-uniform hardware configuration for our analysis of device topology influences.
Four lanes of PCI Express are connected to each NVMe SSD, while sixteen lanes are connected to each GPU.

The host runs Red Hat Enterprise Linux Core OS 4.12 in OpenShift 4.12 (Kubernetes 1.25) on Linux 4.18 with NVIDIA driver version 550.90.07.
Our experiments are conducted in a container built on the Red Hat Universal Base Image 8.6 with CUDA 12.1.
We used safetensors 0.4.3 as the baseline to be compared with fastsafetensors results.

\subsection{Model loading time}

\begin{figure}[t]
\centering
\begin{subfigure}{0.38\linewidth}
\centering
\includegraphics[width=\linewidth]{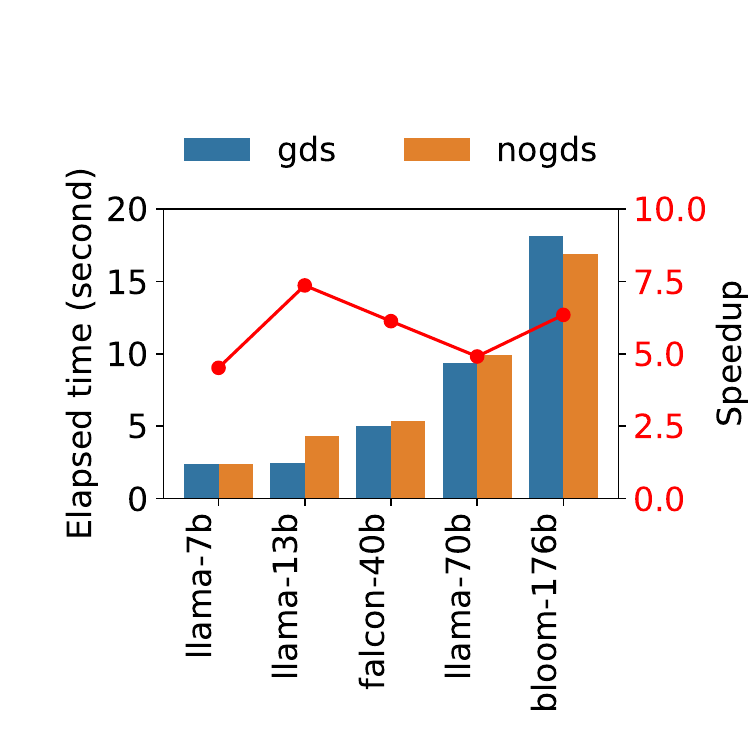}
\caption{Elapsed time.}
\label{fig:fst_elapsed_time}
\end{subfigure}
\hfill
\begin{subfigure}{0.26\linewidth}
\centering
\includegraphics[width=\linewidth]{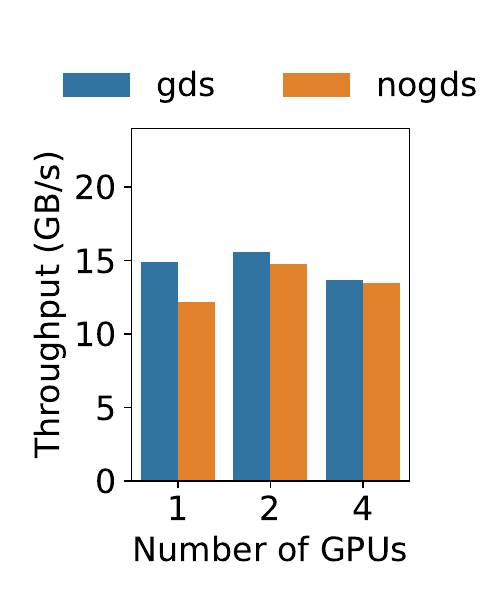}
\caption{Strong scaling of Falcon.}
\label{fig:fst_strong_scaling}
\end{subfigure}
\hfill
\begin{subfigure}{0.26\linewidth}
\centering
\includegraphics[width=\linewidth]{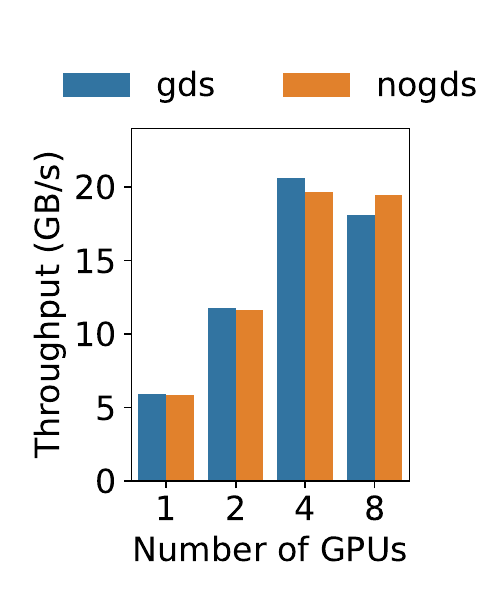}
\caption{Weak scaling of Bloom.}
\label{fig:fst_weak_scaling}
\end{subfigure}
\caption{Performance of fastsafetensors.}
\label{fig:fst_perf}
\end{figure}

The first experiment runs the standalone loader with varying models and numbers of GPUs.
The experiment utilizes a single GPU for Llama-7B and Llama-13B, two GPUs for Falcon, four GPUs for Llama-70B, and eight GPUs for Bloom.
We set CPU affinity to the NUMA node that is connected to the NVMe SSDs, except for Bloom experiments which use both NUMA nodes.
This setup is the same as our preliminary experiments in Section~\ref{sec:motivation:analysis}.

Figure~\ref{fig:fst_elapsed_time} shows the elapsed time of our standalone loader with each configuration and the speedups of fastsafetensors compared to the default safetensors library with GDS enabled.
Fastsafetensors improved both single- and multi-GPU use cases with more than 4.8x speedups.
Bloom showed the lowest speedups due to the lack of NUMA awareness, which we analyze in Section~\ref{sec:exp:add}.
GDS improved the performance of fastsafetensors with every model except Bloom.

Figures~\ref{fig:fst_weak_scaling} and \ref{fig:fst_strong_scaling} show the results of weak and strong scaling.
Fastsafetensors achieved high resource utilization even with a single GPU and demonstrated relatively low strong scaling results (Figure~\ref{fig:fst_strong_scaling}).
%TODO 1-(f): Figure 10b shows that the strong scaling performance of Fastsafetensors is poor. However, the authors do not provide any explanation for this behavior.
The low strong scaling result indicates that the single GPU load was well-optimized and shuffling after deserialization with multiple-GPUs could not be amortized for the data size (78 GB) that is fit to the current capacity of single-GPU memory.
However, the results of weak scaling were linearly improved as the number of GPUs increased from one to four, while the improvement ratio decreased with eight GPUs.

\subsection{Resource utilization}

We also analyze the time-series utilizations of compute resources during TGIS startups and the first inference run.
Overall, the startup time was consistent with the standalone loader results shown in Figure~\ref{fig:fst_elapsed_time}.

\begin{figure}[t]
\centering
\begin{subfigure}{0.48\linewidth}
\centering
\includegraphics[width=\linewidth]{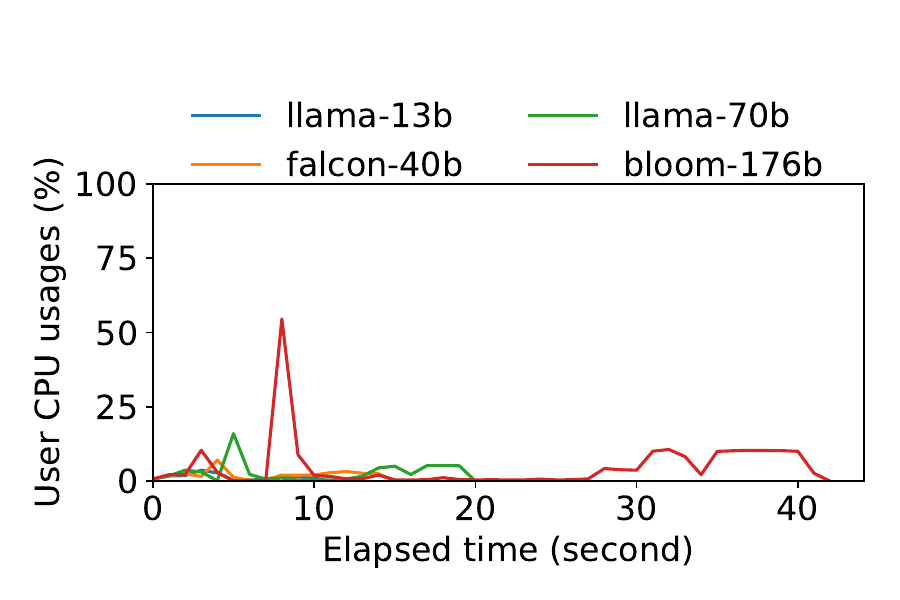}
\caption{User CPU w/ GDS.}
\label{fig:gds_usr}
\end{subfigure}
\hfill
\begin{subfigure}{0.48\linewidth}
\centering
\includegraphics[width=\linewidth]{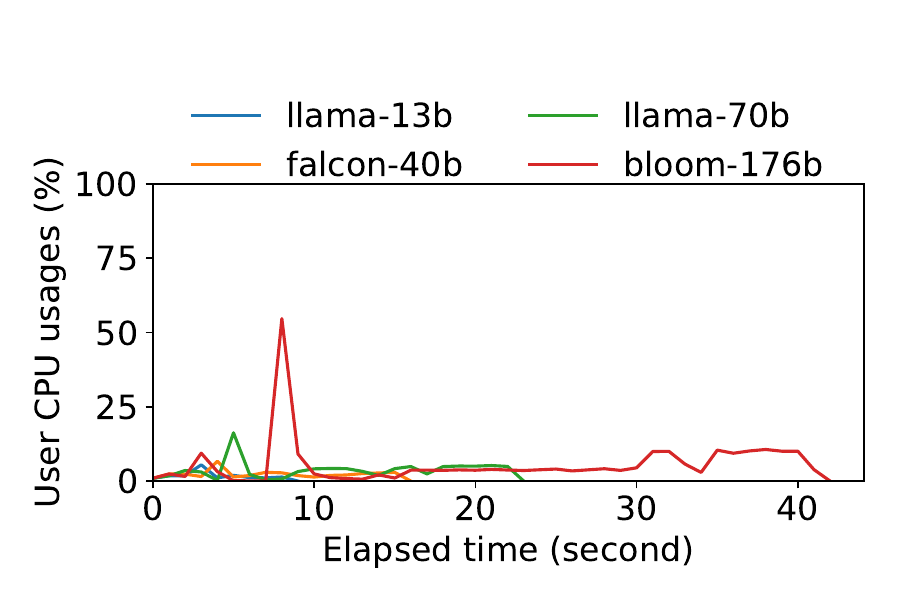}
\caption{User CPU w/o GDS.}
\label{fig:nogds_usr}
\end{subfigure}
\vfill
\begin{subfigure}{0.48\linewidth}
\centering
\includegraphics[width=\linewidth]{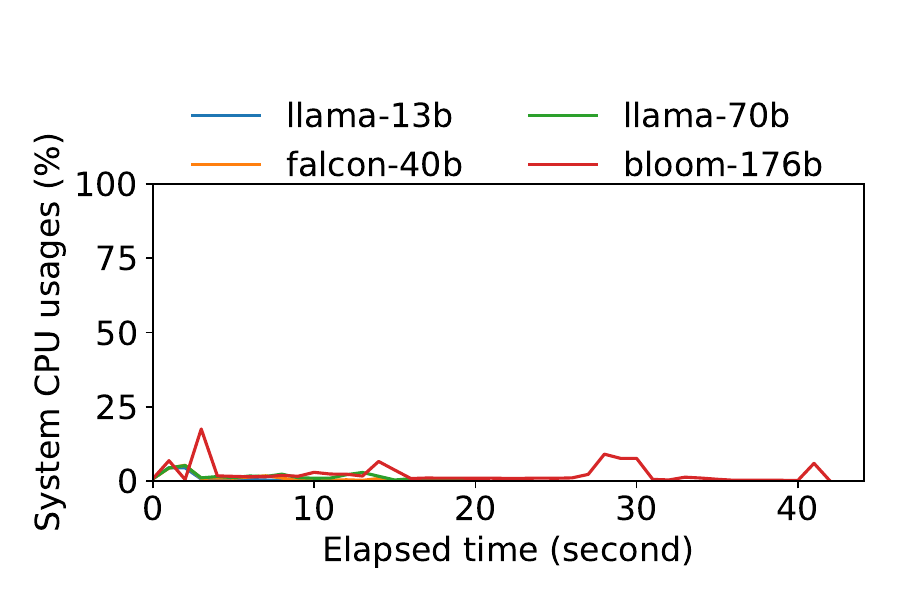}
\caption{Kernel CPU w/ GDS.}
\label{fig:gds_sys}
\end{subfigure}
\hfill
\begin{subfigure}{0.48\linewidth}
\centering
\includegraphics[width=\linewidth]{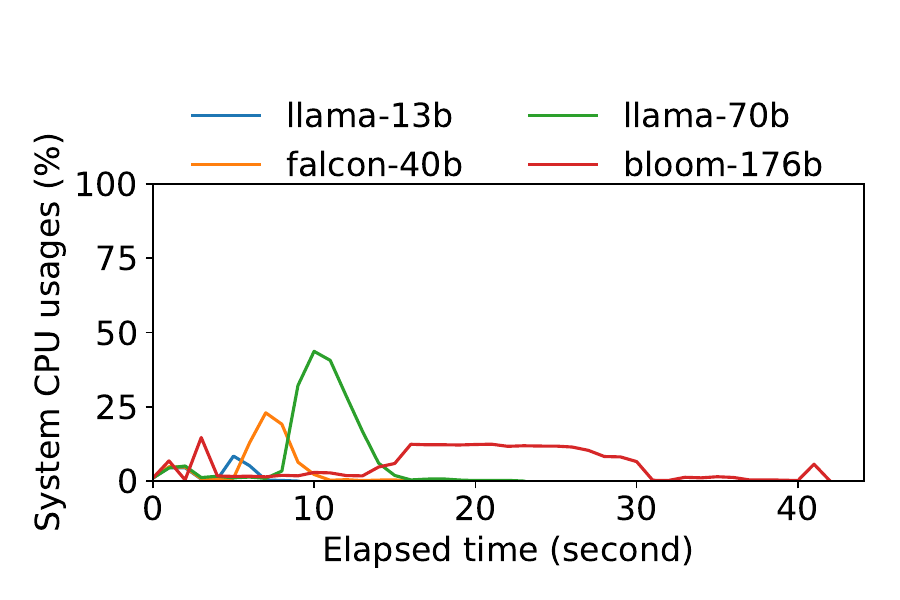}
\caption{Kernel CPU w/o GDS.}
\label{fig:nogds_sys}
\end{subfigure}
\caption{Host CPU usages of fastsafetensors.}
\label{fig:fst_cpu}
\end{figure}

Figures~\ref{fig:gds_usr} and \ref{fig:gds_sys} show that fastsafetensors minimized host CPU usage during model loading.
The overall trends show an initial spike for process initialization, followed by low usage for the remainder of the model loading and inference.
Fastsafetensors mitigated high system CPU usage with its optimized file I/O, and GDS further reduced this usage.
The reduced system CPU usage from GDS implies that it can improve the robustness of workload performance for co-located jobs, which we further analyze in Section~\ref{sec:exp:add}.

\begin{figure}[t]
\centering
\begin{subfigure}{0.48\linewidth}
\centering
\includegraphics[width=\linewidth]{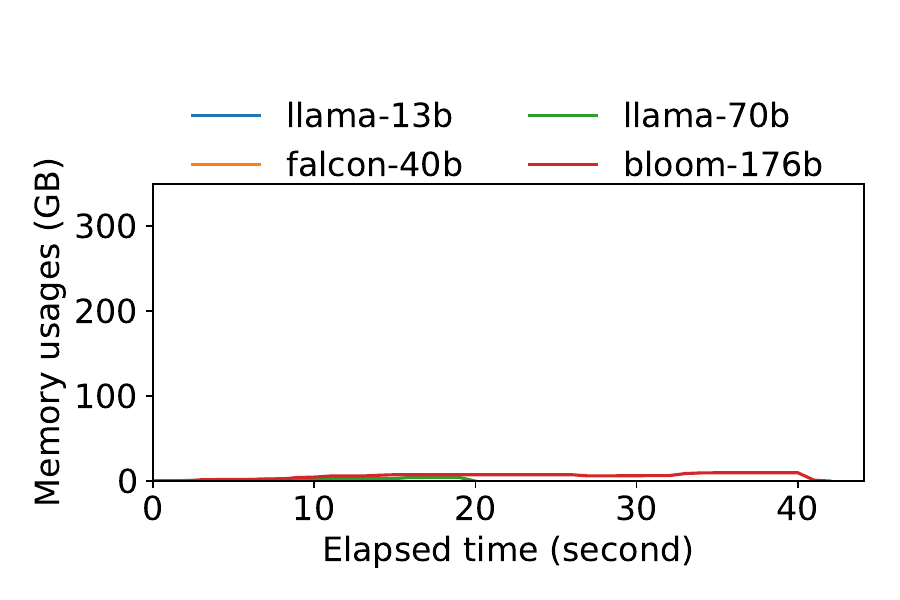}
\caption{User memory w/ GDS.}
\label{fig:gds_used}
\end{subfigure}
\hfill
\begin{subfigure}{0.48\linewidth}
\centering
\includegraphics[width=\linewidth]{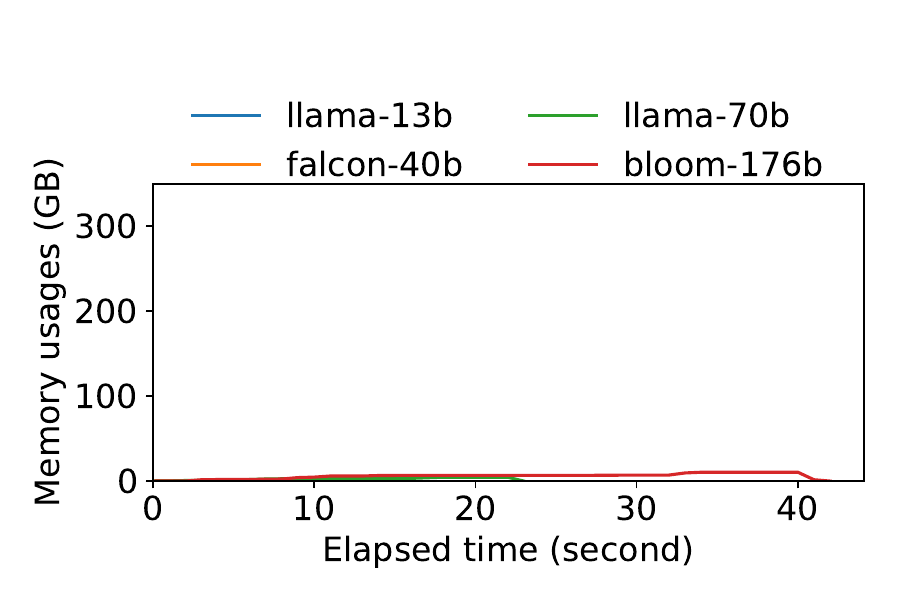}
\caption{User memory w/o GDS.}
\label{fig:nogds_used}
\end{subfigure}
\vfill
\begin{subfigure}{0.48\linewidth}
\centering
\includegraphics[width=\linewidth]{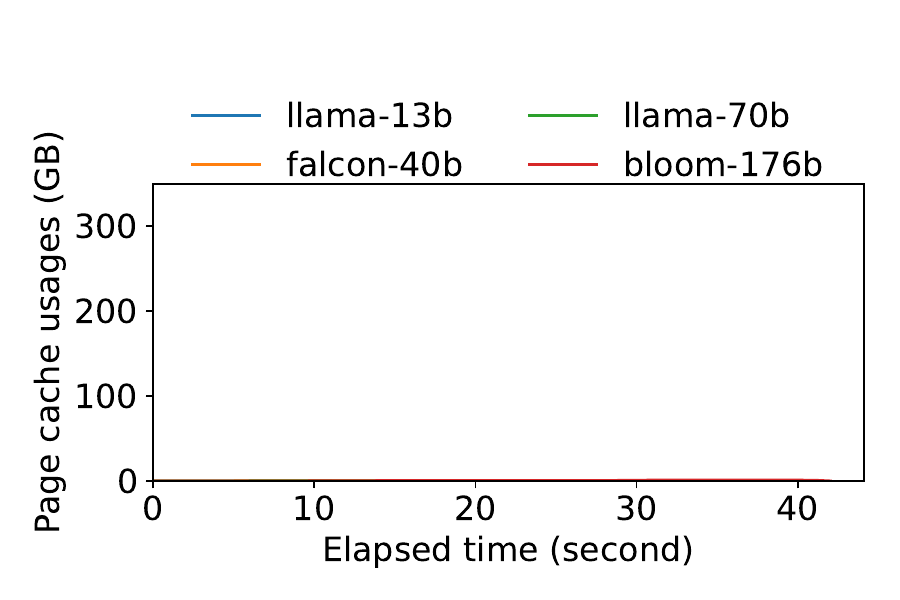}
\caption{Page cache w/ GDS.}
\label{fig:gds_cach}
\end{subfigure}
\hfill
\begin{subfigure}{0.48\linewidth}
\centering
\includegraphics[width=\linewidth]{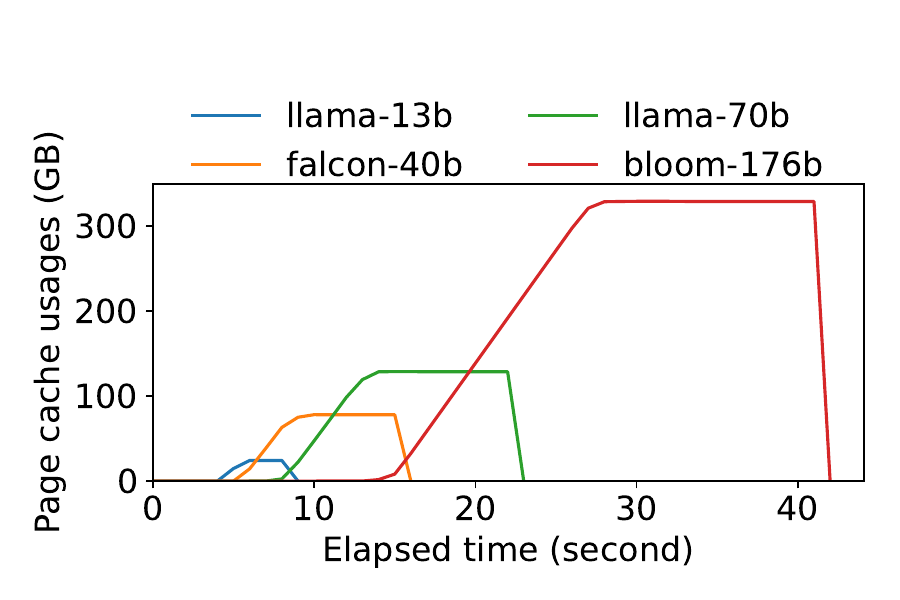}
\caption{Page cache w/o GDS.}
\label{fig:nogds_cach}
\end{subfigure}
\caption{Host memory usages of fastsafetensors.}
\label{fig:fst_mem}
\end{figure}

Figures~\ref{fig:gds_used} and \ref{fig:gds_cach} show that GDS minimized host memory usage by eliminating bounce buffering and page cache usage.
Without GDS, fastsafetensors required notable page cache footprints that were the same size as the model files, regardless of the speedups from our optimizations.
Note that direct I/O is available in Linux to eliminate page cache usage, but we did not enable it due to its limitation regarding supported filesystems, such as tmpfs.

\begin{figure}[t]
\centering
\begin{subfigure}{0.48\linewidth}
\centering
\includegraphics[width=\linewidth]{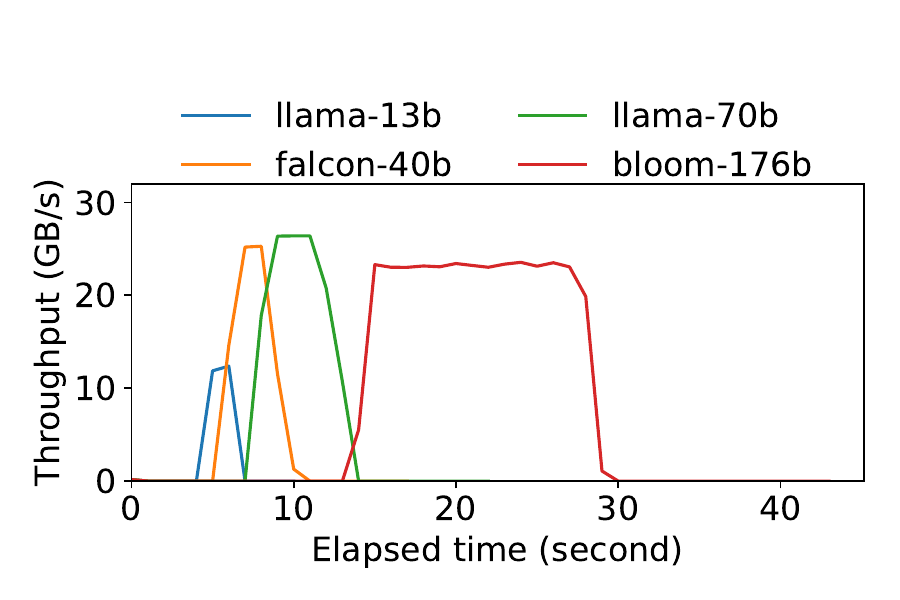}
\caption{Throughput w/ GDS.}
\label{fig:gds_disk}
\end{subfigure}
\hfill
\begin{subfigure}{0.48\linewidth}
\centering
\includegraphics[width=\linewidth]{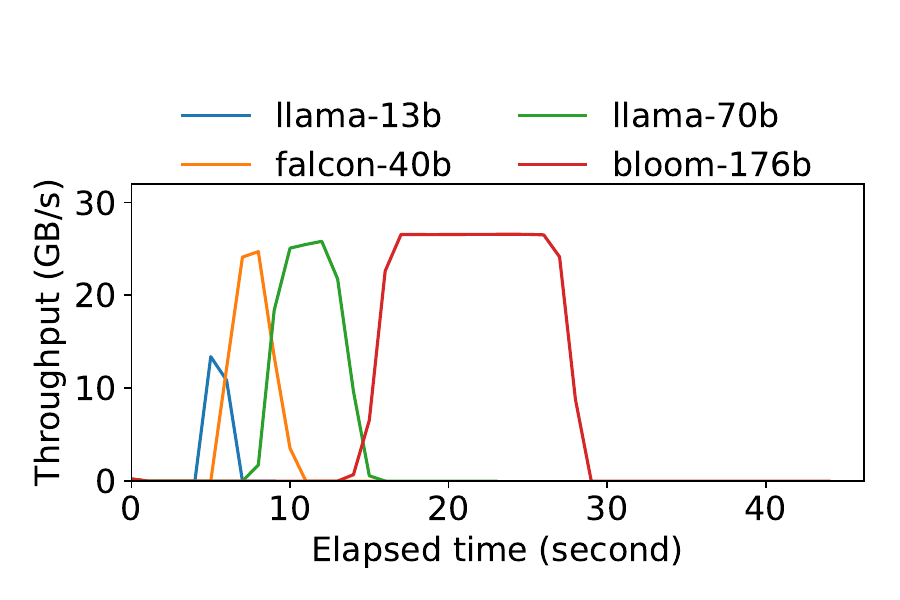}
\caption{Throughput w/o GDS.}
\label{fig:nogds_disk}
\end{subfigure}
\label{fig: fst_disk}
\caption{NVMe SSD read throughput of fastsafetensors.}
\label{fig: fst_mem}
\end{figure}

Figure~\ref{fig:gds_disk} shows that fastsafetensors highly utilized storage throughput, especially when using multiple GPUs.
Llama-70B reached 26.4 GB/s when copying files to four GPUs with GDS, while it slightly decreased to be 25.8 GB/s without GDS.
However, Bloom demonstrated that GDS decreased the maximum storage throughput compared to the results without GDS.
Overall results, both with and without GDS, indicate that the major improvement in storage utilization is derived from our aggregated file copying approach, rather than from GDS itself.

\begin{figure}[t]
\centering
\begin{subfigure}{0.48\linewidth}
\centering
\includegraphics[width=\linewidth]{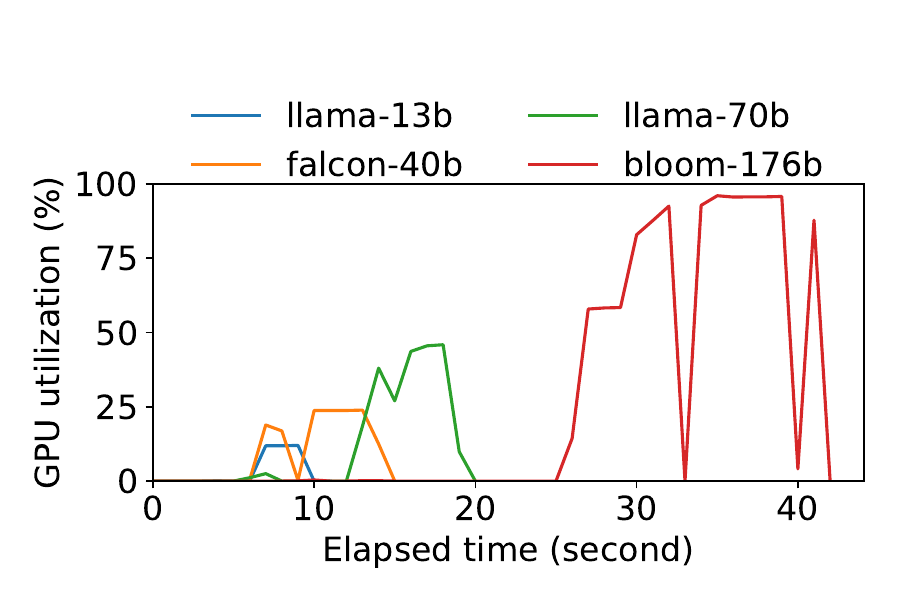}
\caption{GPU utilization w/ GDS.}
\label{fig:gds_util}
\end{subfigure}
\hfill
\begin{subfigure}{0.48\linewidth}
\centering
\includegraphics[width=\linewidth]{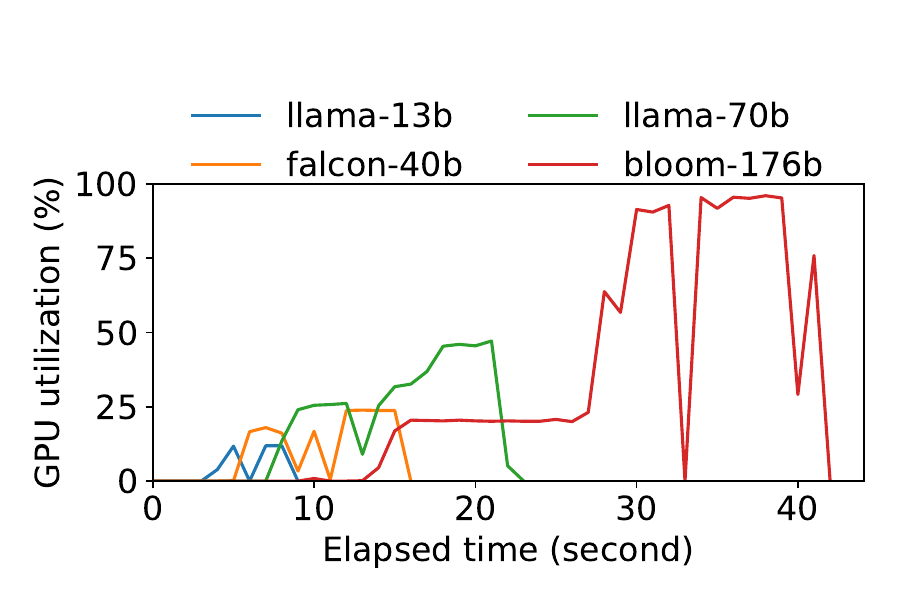}
\caption{GPU utilization w/o GDS.}
\label{fig:nogds_util}
\end{subfigure}
\hfill
\begin{subfigure}{0.48\linewidth}
\centering
\includegraphics[width=\linewidth]{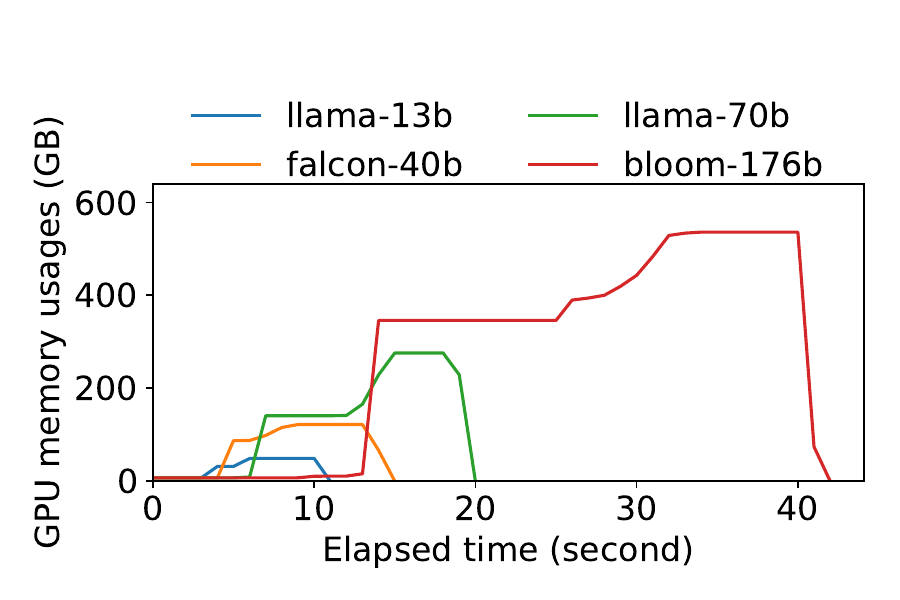}
\caption{GPU memory w/ GDS.}
\label{fig:gds_gpumem}
\end{subfigure}
\hfill
\begin{subfigure}{0.48\linewidth}
\centering
\includegraphics[width=\linewidth]{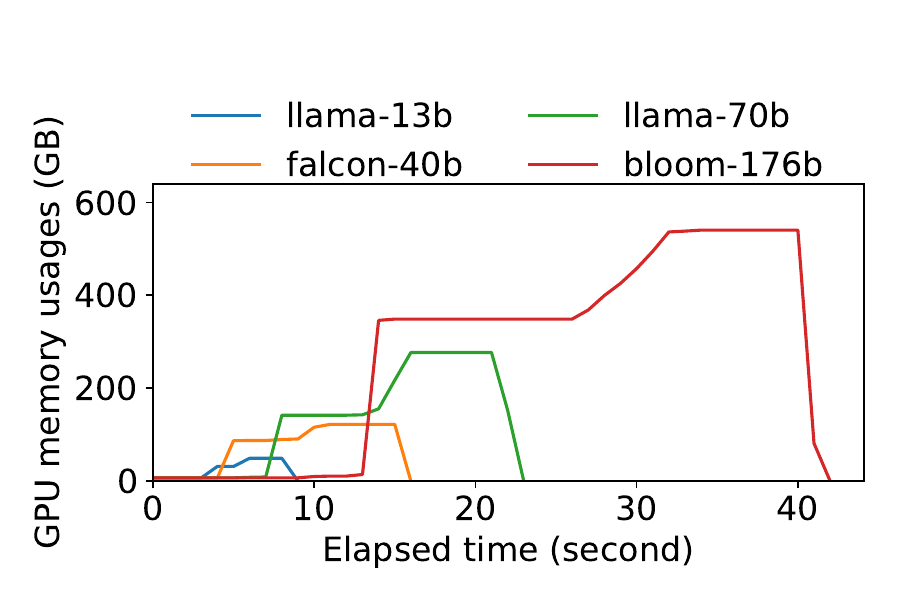}
\caption{GPU memory w/o GDS.}
\label{fig:nogds_gpumem}
\end{subfigure}
\vfill
\begin{subfigure}{0.48\linewidth}
\centering
\includegraphics[width=\linewidth]{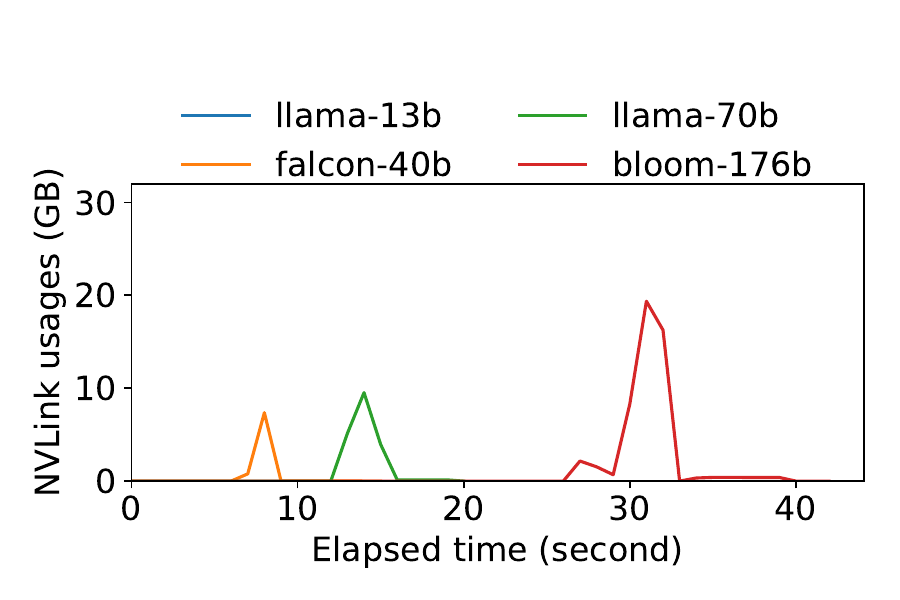}
\caption{NVLink w/ GDS.}
\label{fig:gds_nvl}
\end{subfigure}
\hfill
\begin{subfigure}{0.48\linewidth}
\centering
\includegraphics[width=\linewidth]{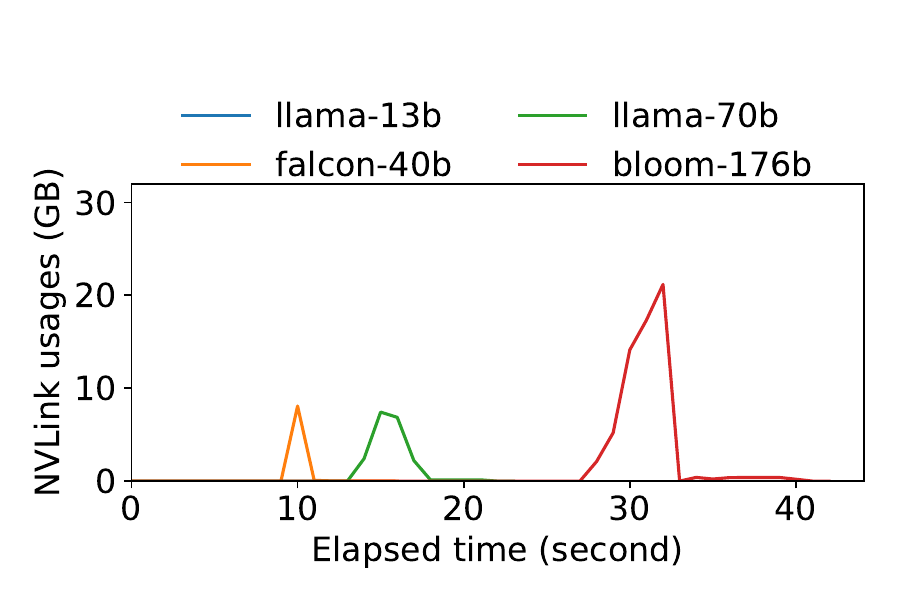}
\caption{NVLink w/o GDS.}
\label{fig:nogds_nvl}
\end{subfigure}
\hfill
\caption{GPU resource usages of fastsafetensors.}
\label{fig:gds_src_usages}
\end{figure}

Figure~\ref{fig:gds_src_usages} shows that fastsafetensors highly utilize GPU resources during model loading.
Figures~\ref{fig:gds_gpumem} and \ref{fig:nogds_gpumem} allow us to easily understand the behavior of fastsafetensors.
Fastsafetensors first calculates the amounts of file reads to allocate GPU memory, unlike the default safetensors library, which allocates tensors one by one.
After memory allocation, fastsafetensors triggers file copies to GPU memory and then partitions tensors with shuffling for multi-GPU workloads.
The experiments also included inference computation to increase GPU memory usage for iterative decoder processing to generate 128 tokens with 263 input tokens.

Shuffling requires additional memory buffering at GPUs, but it was not notably high since it kept reusing GPU memory with the Torch memory allocator and its caching system.
% TODO 1-(g): Shuffling requires additional memory buffering on the GPUs; however, the authors do not provide concrete numbers to determine whether the extra memory usage is significant.
We configured the system to allocate a 10 GB GPU buffer to directly deserialize a safetensors file for each GPU, and shuffle them among GPUs on newly-allocated tensors.
The 10-GB GPU buffers are re-used in the shuffle phase or inference computation immediately after the deserialization phase.
Shuffling also resulted in high NVLink utilization, as described in Figures~\ref{fig:gds_nvl} and \ref{fig:nogds_nvl}.
Figures~\ref{fig:gds_util} and \ref{fig:nogds_util} show that shuffling also caused high GPU utilization.
However, they also indicate that GDS reduced GPU utilization during tensor copies.

In summary, fastsafetensors dramatically improved resource efficiency during model loading and minimized elapsed time with optimized file I/O and shuffling.
GDS further improved resource efficiency, but it was a relatively minor factor in its improvement under simple situations, such as a single inference run without co-located jobs on a node.
%Therefore, we argue that fastsafetensors without GDS has enough values regardless of its compromised performance since GDS has strict tradeoffs in practical deployments and other GPU features as described in Section~\ref{sec:discussion}.

\subsection{Extended Performance Evaluation}
\label{sec:exp:add}

\begin{figure}[t]
\centering
\begin{subfigure}{0.30\linewidth}
\centering
\includegraphics[width=\linewidth]{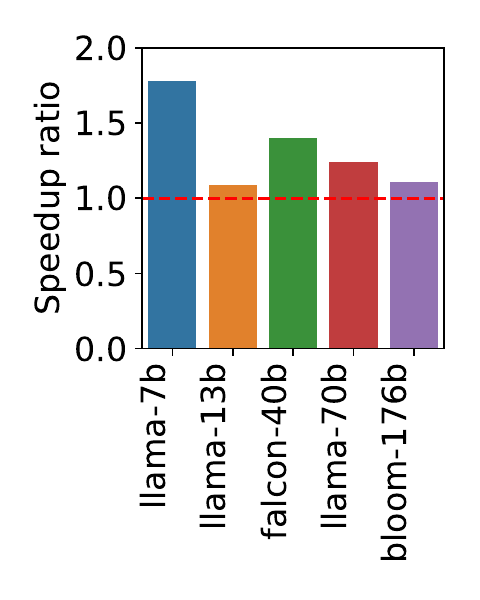}
\caption{DRAM-backed storage.}
\label{fig:nvme_vs_tmpfs}
\end{subfigure}
\hfill
\begin{subfigure}{0.32\linewidth}
\centering
\includegraphics[width=\linewidth]{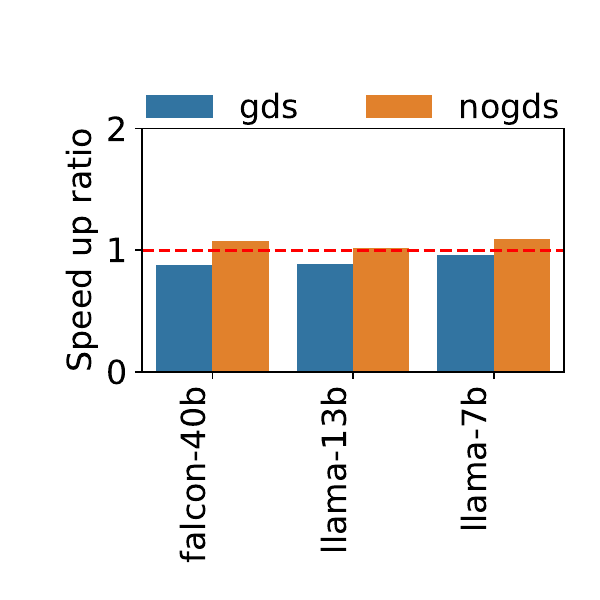}
\caption{NUMA ignorance.}
\label{fig:numa}
\end{subfigure}
\hfill
\begin{subfigure}{0.32\linewidth}
\centering
\includegraphics[width=\linewidth]{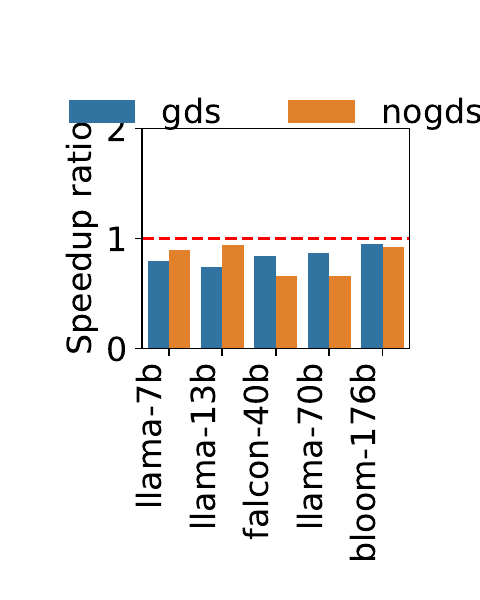}
\caption{Background memcpy.}
\label{fig:fill_l2}
\end{subfigure}
\caption{Performance speedup under different situations: 
Each figure shows the relative speedup from the workloads under four NVMe SSDs with NUMA awareness and no co-located jobs.
Results higher than 1.0 indicate speedups, while results lower than 1.0 indicate slowdowns due to the changed situation.
}
\label{fig:add_exp}
\end{figure}

Section~\ref{sec:exp:overall} characterized the behavior of fastsafetensors under a simple deployment, but it still leaves questions about other practical situations.
In this section, we analyze performance variations of fastsafetensors under 1) files cached on DRAM, 2) device (NUMA) toplogy ignorance, and 3) co-located jobs.
For 1), we set up DRAM files on tmpfs and test the same model loading as in the prior experiments.
For 2), we explicitly bind an available GPU to a different NUMA node from the four NVMe SSDs and run single-GPU model loading on Llama-7B, Llama-13B, and Falcon-40B.
We experimented with synthetic co-located jobs that involve concurrent memory copies (building a 160-MB buffer and sequentially filling a 64-bit number infinitely) on every CPU core in the same NUMA node as the NVMe SSDs while the model loading runs on the node.

Figure~\ref{fig:nvme_vs_tmpfs} shows relative speedups when using tmpfs with GDS disabled instead of using four NVMe SSDs with GDS enabled.
We observed higher speedups for Llama-7B but relatively low benefits for larger model loading with multiple GPUs.
Overall improvements are lower than the speed difference between DRAM and NVMe SSDs because PCI express bus bandwidth (32 GB/s for 16 lanes in Gen 4) becomes the key factor in determining the maximum throughput for our experiment.
This result implies that memory extension with PCI Express is preferable in some cases in terms of cost efficiency for model loading compared to reserving or increasing DRAM capacity.

Figure~\ref{fig:numa} shows speedups or slowdowns with NUMA ignorance.
NUMA ignorance degraded the performance of model loading with GDS enabled.
In contrast, we observed performance improvements with GDS disabled since it can utilize the CPU cache of both NUMA nodes during file reads and copies to GPU memory.
These results indicate that system administrators or job schedulers should tune device topology for GDS, and disabling GDS can be a viable choice when tuning is not easy.

Figure~\ref{fig:fill_l2} shows slowdowns with concrreunt memory copies.
All models demonstrated that enabling GDS improved performance robustness against co-located jobs.
These results are consistent with our resource analysis in the previous section, which showed that GDS minimized host CPU and memory usage.
Thus, GDS has relatively high benefits under high workload density on a machine, where many jobs are co-located to improve cost efficiency.

\subsection{vLLM Extension}

%https://github.com/vllm-project/vllm/pull/10647

\begin{table}
\begin{tabular}{lcccc}
\toprule
Model & \# of GPUs & GPU & Baseline & Fastsafetensors \\
\midrule
Llama-2-7b-hf & 1 & A100 & 7.29 & 3.67 \\
Llama-2-13b-hf & 1 & A100 & 16.04 & 6.88 \\
Llama-2-13b-hf & 2 & A100 & 14.35 & 6.02 \\
Llama-2-13b-hf & 4 & L40S & 12.39 & 4.74 \\
\bottomrule
\end{tabular}
\caption{vLLM startup time (seconds).}
\label{fig:vllm}
\end{table}

We integrated fastsafetensors with the vLLM~\cite{vLLM} and evaluated its performance.
Our testing was conducted with the upstream version of vLLM as of March 3, 2025.
The implementation involved a modification of 42 lines of Python code, where the generator code that loads safetensors files was replaced with calls to the fastsafetensors API.

The performance results of this integration, shown in Table~\ref{fig:vllm}, were obtained by testing various models with different numbers of GPUs.
The models tested include Llama-2-7b-hf and Llama-2-13b-hf, using both A100 and L40S GPU configurations.
As shown in the table, fastsafetensors consistently outperformed the baseline in all configurations, reducing the startup time significantly.
The improvements were observed across different models and GPU setups, including both single-GPU (A100) and multi-GPU (A100 and L40S) configurations.

These results demonstrate the effectiveness of fastsafetensors in reducing the model loading times within vLLM.
Our future work is testing additional models and further optimizing the integration to maximize performance across a wider range of configurations.

\section{Related work}
\label{sec:related}

This work improves the deserialization method of the widely used safetensors format for pre-trained models.
In this section, we discuss existing work in the related fields of data serialization and deserialization.

Various formats have been explored to represent model weights for machine learning, including those for LLMs.
Each format employs different deserialization methods.
Safetensors~\cite{Safetensors}, the focus of this work, is designed to be safer and faster compared to Python Pickle, as explained in Section~\ref{sec:motivation}.
Tensorizer~\cite{Tensorizer} focuses on speeding up access, especially for S3 and HTTP access.
However, it does not include the application of GDS or tensor preprocessing offloading, leaving room for further optimization.

ServerlessLLM~\cite{ServerlessLLM} achieves fast startup for inference services by caching to model storage, converting to efficient formats, and scheduling with locality considerations.
While ServerlessLLM optimizes the startup process rather than the loading efficiency itself, incorporating fastsafetensors could further reduce startup times.

In the context of LLM training workloads, ZeroInfinity~\cite{ZeroInfinity} adds NVMe to offload tensors, effectively overlapping device-to-device communication to train large models with limited GPU memory.
Although direct offloading between GPU and NVMe is not included, further efficiency could be achieved by reusing GDS and aggregated tensor deserialization, similar to this work.

In data analytics workloads like Spark, additional metadata is used to improve data read efficiency.
Column Cache~\cite{ColumnCache-BigData18} utilizes Parquet format and execution plan information in Spark SQL for efficient data reading.
TeraHeap~\cite{TeraHeap-ASPLOS23} reduces overhead by using garbage collection hints when serializing and deserializing objects temporarily in data analytics frameworks.
Nozawa et al.~\cite{InMemArrShare-BigData23} use format information to optimize sharing of Apache Arrow files across multiple processes.
%Fastsafetensors similarly utilizes tensor file offsets to delay tensor instantiation after file copies.
Fastsafetensors can also be further improved by utilizing information from execution plans generated with tools like Torch Dynamo or Python runtime information.

\section{Discussion}
\label{sec:discussion}

As demonstrated in our experiments, fastsafetensors significantly improved model loading performance and efficiency compared to existing safetensors library.
However, an open question remains about when to enable GDS.
While GDS generally provides performance benefits and even can enhance relative speeds under co-located jobs,
it also imposes strict constraints related to device toplogy to avoid performance degradation.
From a system administration perspective, GDS has strict requirements for software deployment such as the installation of kernel module and user-level library, along with the replacement of the NVMe kernel module.
This means that rebooting the entire machine may be required when installing the required kernel modules, particularly if system services are actively using the NVMe SSDs.
Therefore, an additional cluster and/or process scheduling system that can accommodate these requirements is preferable for GDS applications.

Fastsafetensors relies on the DLPack specification to cover its supported data types.
For example, the current Pytorch does not enable us to directly deserialize new data types such as eight-bit floating point~\cite{FP8-Arxiv} with DLPack.
This is not a technical limitation, but this led to a lack of timely support for FP8 quantization~\cite{Quantization-Arxiv,FP8Quantization-Arxiv}.

This work does not mention other accelerators besides GPUs.
However, we believe that the performance problem fastsafetensors addresses is derived from the design issue of tensor object management, which is common across accelerators.

% TODO 2-2: Any comment on whether similar optimizations can be applied when writing checkpoints / safetensors? I agree the problem of deserialization is much more important as it has a significant impact on model loading time. But, even if only from an academic standpoint, it will be useful to consider what is possible when writing tensors to disk.
Our approach of batched tensor instantiation could potentially be adapted for efficient serialization from device memory to storage.
One of the challenges is that current device memory management does not ensure locating model parameters in a contiguous region, which is required for efficient data copies.

\section{Conclusion}
\label{sec:conclusion}

In this work, we proposed fastsafetensors to address the inefficient designs of the current safetensors deserializer.
Fastsafetensors significantly improved resource utilization through aggregated tensor deserialization, shuffling, and GDS.
As a result, a substantial reduction in loading time (from 4.8x to 7.5x) is demonstrated compared to the current deserializer.
The fastsafetensors code is available at \url{https://github.com/foundation-model-stack/fastsafetensors}.

\bibliography{papers}

\begin{thebibliography}{10}
\providecommand{\url}[1]{#1}
\csname url@samestyle\endcsname
\providecommand{\newblock}{\relax}
\providecommand{\bibinfo}[2]{#2}
\providecommand{\BIBentrySTDinterwordspacing}{\spaceskip=0pt\relax}
\providecommand{\BIBentryALTinterwordstretchfactor}{4}
\providecommand{\BIBentryALTinterwordspacing}{\spaceskip=\fontdimen2\font plus
\BIBentryALTinterwordstretchfactor\fontdimen3\font minus
  \fontdimen4\font\relax}
\providecommand{\BIBforeignlanguage}[2]{{%
\expandafter\ifx\csname l@#1\endcsname\relax
\typeout{** WARNING: IEEEtran.bst: No hyphenation pattern has been}%
\typeout{** loaded for the language `#1'. Using the pattern for}%
\typeout{** the default language instead.}%
\else
\language=\csname l@#1\endcsname
\fi
#2}}
\providecommand{\BIBdecl}{\relax}
\BIBdecl

\bibitem{ChatGPT}
\BIBentryALTinterwordspacing
{OpenAI}, ``{Introducing ChatGPT},'' 2022. [Online]. Available:
  \url{https://openai.com/index/chatgpt/}
\BIBentrySTDinterwordspacing

\bibitem{Gemini}
\BIBentryALTinterwordspacing
S.~Pichai and D.~Hassabis, ``{Introducing Gemini: our largest and most capable
  AI model},'' 2023. [Online]. Available:
  \url{https://blog.google/technology/ai/google-gemini-ai/}
\BIBentrySTDinterwordspacing

\bibitem{GraniteCode}
\BIBentryALTinterwordspacing
M.~Mishra, M.~Stallone, G.~Zhang, Y.~Shen, A.~Prasad, A.~M. Soria, M.~Merler,
  P.~Selvam, S.~Surendran, S.~Singh, M.~Sethi, X.-H. Dang, P.~Li, K.-L. Wu,
  S.~Zawad, A.~Coleman, M.~White, M.~Lewis, R.~Pavuluri, Y.~Koyfman,
  B.~Lublinsky, M.~de~Bayser, I.~Abdelaziz, K.~Basu, M.~Agarwal, Y.~Zhou,
  C.~Johnson, A.~Goyal, H.~Patel, Y.~Shah, P.~Zerfos, H.~Ludwig, A.~Munawar,
  M.~Crouse, P.~Kapanipathi, S.~Salaria, B.~Calio, S.~Wen, S.~Seelam,
  B.~Belgodere, C.~Fonseca, A.~Singhee, N.~Desai, D.~D. Cox, R.~Puri, and
  R.~Panda, ``{Granite Code Models: A Family of Open Foundation Models for Code
  Intelligence},'' 2024. [Online]. Available:
  \url{https://arxiv.org/abs/2405.04324}
\BIBentrySTDinterwordspacing

\bibitem{CompoundAISystem}
\BIBentryALTinterwordspacing
M.~Zaharia, O.~Khattab, L.~Chen, J.~Q. Davis, H.~Miller, C.~Potts, J.~Zou,
  M.~Carbin, J.~Frankle, N.~Rao, and A.~Ghodsi, ``{The Shift from Models to
  Compound AI Systems},'' 2024. [Online]. Available:
  \url{https://bair.berkeley.edu/blog/2024/02/18/compound-ai-systems/}
\BIBentrySTDinterwordspacing

\bibitem{FlashAttention-NeurIPS22}
T.~Dao, D.~Y. Fu, S.~Ermon, A.~Rudra, and C.~R{\'e}, ``Flash{A}ttention: Fast
  and memory-efficient exact attention with {IO}-awareness,'' in \emph{Advances
  in Neural Information Processing Systems (NeurIPS '22)}, 2022.

\bibitem{FlashAttention2-ICLR24}
T.~Dao, ``Flash{A}ttention-2: Faster attention with better parallelism and work
  partitioning,'' in \emph{International Conference on Learning Representations
  (ICLR '24)}, 2024.

\bibitem{PagedAttention-SOSP23}
W.~Kwon, Z.~Li, S.~Zhuang, Y.~Sheng, L.~Zheng, C.~H. Yu, J.~Gonzalez, H.~Zhang,
  and I.~Stoica, ``{Efficient Memory Management for Large Language Model
  Serving with PagedAttention},'' in \emph{{Proceedings of the 29th Symposium
  on Operating Systems Principles (SOSP '23)}}, 2023, p. 611–626.

\bibitem{Orca-OSDI22}
G.-I. Yu, J.~S. Jeong, G.-W. Kim, S.~Kim, and B.-G. Chun, ``{Orca: A
  Distributed Serving System for {Transformer-Based} Generative Models},'' in
  \emph{{16th USENIX Symposium on Operating Systems Design and Implementation
  (OSDI '22)}}, 2022, pp. 521--538.

\bibitem{SpeculativeDecoding}
\BIBentryALTinterwordspacing
D.~Wertheimer, J.~Rosenkranz, T.~Parnell, S.~Suneja, P.~Ranganathan, R.~Ganti,
  and M.~Srivatsa, ``{Accelerating Production LLMs with Combined
  Token/Embedding Speculators},'' 2024. [Online]. Available:
  \url{https://arxiv.org/abs/2404.19124}
\BIBentrySTDinterwordspacing

\bibitem{DistServe}
\BIBentryALTinterwordspacing
Y.~Zhong, S.~Liu, J.~Chen, J.~Hu, Y.~Zhu, X.~Liu, X.~Jin, and H.~Zhang,
  ``{DistServe: Disaggregating Prefill and Decoding for Goodput-optimized Large
  Language Model Serving},'' 2024. [Online]. Available:
  \url{https://arxiv.org/abs/2401.09670}
\BIBentrySTDinterwordspacing

\bibitem{Sarathi-OSDI24}
A.~Agrawal, N.~Kedia, A.~Panwar, J.~Mohan, N.~Kwatra, B.~S. Gulavani,
  A.~Tumanov, and R.~Ramjee, ``Taming throughput-latency tradeoff in llm
  inference with sarathi-serve,'' \emph{Proceedings of 18th USENIX Symposium on
  Operating Systems Design and Implementation, (OSDI '24)}, 2024.

\bibitem{Tensorizer}
\BIBentryALTinterwordspacing
N.~Pratt and R.~Talari, ``{Decrease PyTorch Model Load Times with CoreWeave’s
  Tensorizer},'' 2024. [Online]. Available:
  \url{https://www.coreweave.com/blog/coreweaves-tensorizer-decrease-pytorch-model-load-times}
\BIBentrySTDinterwordspacing

\bibitem{ServerlessLLM}
\BIBentryALTinterwordspacing
Y.~Fu, L.~Xue, Y.~Huang, A.-O. Brabete, D.~Ustiugov, Y.~Patel, and L.~Mai,
  ``{ServerlessLLM: Locality-Enhanced Serverless Inference for Large Language
  Models},'' 2024. [Online]. Available: \url{https://arxiv.org/abs/2401.14351}
\BIBentrySTDinterwordspacing

\bibitem{Megatron-Arxiv}
\BIBentryALTinterwordspacing
D.~Narayanan, M.~Shoeybi, J.~Casper, P.~LeGresley, M.~Patwary, V.~A.
  Korthikanti, D.~Vainbrand, P.~Kashinkunti, J.~Bernauer, B.~Catanzaro,
  A.~Phanishayee, and M.~Zaharia, ``{Efficient Large-Scale Language Model
  Training on GPU Clusters Using Megatron-LM},'' 2021. [Online]. Available:
  \url{https://arxiv.org/abs/2104.04473}
\BIBentrySTDinterwordspacing

\bibitem{Safetensors}
\BIBentryALTinterwordspacing
{Hugging Face}, ``{Safetensors},'' 2024. [Online]. Available:
  \url{https://huggingface.co/docs/safetensors/}
\BIBentrySTDinterwordspacing

\bibitem{HuggingFace-Models}
\BIBentryALTinterwordspacing
``{Models -- Hugging Face},'' 2025. [Online]. Available:
  \url{https://huggingface.co/models?library=safetensors}
\BIBentrySTDinterwordspacing

\bibitem{Pickle}
\BIBentryALTinterwordspacing
{Python Software Foundation}, ``{pickle --- Python object Serialization},''
  2024. [Online]. Available:
  \url{https://docs.python.org/3/library/pickle.html}
\BIBentrySTDinterwordspacing

\bibitem{Pytorch}
\BIBentryALTinterwordspacing
{The Linux Foundation}, ``{Pytorch},'' 2024. [Online]. Available:
  \url{https://pytorch.org/}
\BIBentrySTDinterwordspacing

\bibitem{TensorFlow-OSDI16}
M.~Abadi, P.~Barham, J.~Chen, Z.~Chen, A.~Davis, J.~Dean, M.~Devin,
  S.~Ghemawat, G.~Irving, M.~Isard, M.~Kudlur, J.~Levenberg, R.~Monga,
  S.~Moore, D.~G. Murray, B.~Steiner, P.~Tucker, V.~Vasudevan, P.~Warden,
  M.~Wicke, Y.~Yu, and X.~Zheng, ``{TensorFlow: A System for Large-Scale
  Machine Learning},'' in \emph{12th USENIX Symposium on Operating Systems
  Design and Implementation (OSDI '16)}, 2016, pp. 265--283.

\bibitem{ZeroInfinity}
\BIBentryALTinterwordspacing
S.~Rajbhandari, O.~Ruwase, J.~Rasley, S.~Smith, and Y.~He, ``{ZeRO-Infinity:
  Breaking the GPU Memory Wall for Extreme Scale Deep Learning},'' 2021.
  [Online]. Available: \url{https://arxiv.org/abs/2104.07857}
\BIBentrySTDinterwordspacing

\bibitem{FDSP-Arxiv}
\BIBentryALTinterwordspacing
Y.~Zhao, A.~Gu, R.~Varma, L.~Luo, C.-C. Huang, M.~Xu, L.~Wright,
  H.~Shojanazeri, M.~Ott, S.~Shleifer, A.~Desmaison, C.~Balioglu, P.~Damania,
  B.~Nguyen, G.~Chauhan, Y.~Hao, A.~Mathews, and S.~Li, ``{PyTorch FSDP:
  Experiences on Scaling Fully Sharded Data Parallel},'' 2023. [Online].
  Available: \url{https://arxiv.org/abs/2304.11277}
\BIBentrySTDinterwordspacing

\bibitem{ByteCheckpoint-Arxiv}
\BIBentryALTinterwordspacing
B.~Wan, M.~Han, Y.~Sheng, Y.~Peng, H.~Lin, M.~Zhang, Z.~Lai, M.~Yu, J.~Zhang,
  Z.~Song, X.~Liu, and C.~Wu, ``{ByteCheckpoint: A Unified Checkpointing System
  for Large Foundation Model Development},'' 2024. [Online]. Available:
  \url{https://arxiv.org/abs/2407.20143}
\BIBentrySTDinterwordspacing

\bibitem{DLPack}
\BIBentryALTinterwordspacing
{DLPack contributors}, ``{Welcome to DLPack's documentation! --- DLPack 0.6.0
  documentation},'' 2022. [Online]. Available:
  \url{https://dmlc.github.io/dlpack/latest/}
\BIBentrySTDinterwordspacing

\bibitem{GDS}
\BIBentryALTinterwordspacing
NVIDIA, ``{NVIDIA Magnum IO GPUDirect Storage Design Guide},'' 2025. [Online].
  Available:
  \url{https://docs.nvidia.com/gpudirect-storage/design-guide/index.html}
\BIBentrySTDinterwordspacing

\bibitem{Llama-Arxiv}
\BIBentryALTinterwordspacing
H.~Touvron, L.~Martin, K.~Stone, P.~Albert, A.~Almahairi, Y.~Babaei,
  N.~Bashlykov, S.~Batra, P.~Bhargava, S.~Bhosale, D.~Bikel, L.~Blecher, C.~C.
  Ferrer, M.~Chen, G.~Cucurull, D.~Esiobu, J.~Fernandes, J.~Fu, W.~Fu,
  B.~Fuller, C.~Gao, V.~Goswami, N.~Goyal, A.~Hartshorn, S.~Hosseini, R.~Hou,
  H.~Inan, M.~Kardas, V.~Kerkez, M.~Khabsa, I.~Kloumann, A.~Korenev, P.~S.
  Koura, M.-A. Lachaux, T.~Lavril, J.~Lee, D.~Liskovich, Y.~Lu, Y.~Mao,
  X.~Martinet, T.~Mihaylov, P.~Mishra, I.~Molybog, Y.~Nie, A.~Poulton,
  J.~Reizenstein, R.~Rungta, K.~Saladi, A.~Schelten, R.~Silva, E.~M. Smith,
  R.~Subramanian, X.~E. Tan, B.~Tang, R.~Taylor, A.~Williams, J.~X. Kuan,
  P.~Xu, Z.~Yan, I.~Zarov, Y.~Zhang, A.~Fan, M.~Kambadur, S.~Narang,
  A.~Rodriguez, R.~Stojnic, S.~Edunov, and T.~Scialom, ``{Llama 2: Open
  Foundation and Fine-Tuned Chat Models},'' 2023. [Online]. Available:
  \url{https://arxiv.org/abs/2307.09288}
\BIBentrySTDinterwordspacing

\bibitem{Falcon-Arxiv}
\BIBentryALTinterwordspacing
E.~Almazrouei, H.~Alobeidli, A.~Alshamsi, A.~Cappelli, R.~Cojocaru, M.~Debbah,
  {\'E}.~Goffinet, D.~Hesslow, J.~Launay, Q.~Malartic, D.~Mazzotta, B.~Noune,
  B.~Pannier, and G.~Penedo, ``{The Falcon Series of Open Language Models},''
  2023. [Online]. Available: \url{https://arxiv.org/abs/2311.16867}
\BIBentrySTDinterwordspacing

\bibitem{Bloom-Arxiv}
\BIBentryALTinterwordspacing
T.~L. Scao, A.~Fan, C.~Akiki, E.~Pavlick, S.~Ilić, D.~Hesslow, R.~Castagné,
  A.~S. Luccioni, F.~Yvon, M.~Gallé, J.~Tow, A.~M. Rush, S.~Biderman,
  A.~Webson, P.~S. Ammanamanchi, T.~Wang, B.~Sagot, N.~Muennighoff, A.~V. del
  Moral, O.~Ruwase, R.~Bawden, S.~Bekman, A.~McMillan-Major, I.~Beltagy,
  H.~Nguyen, L.~Saulnier, S.~Tan, P.~O. Suarez, V.~Sanh, H.~Laurençon,
  Y.~Jernite, J.~Launay, M.~Mitchell, C.~Raffel, A.~Gokaslan, A.~Simhi,
  A.~Soroa, A.~F. Aji, A.~Alfassy, A.~Rogers, A.~K. Nitzav, C.~Xu, C.~Mou,
  C.~Emezue, C.~Klamm, C.~Leong, D.~van Strien, D.~I. Adelani, D.~Radev, E.~G.
  Ponferrada, E.~Levkovizh, E.~Kim, E.~B. Natan, F.~D. Toni, G.~Dupont,
  G.~Kruszewski, G.~Pistilli, H.~Elsahar, H.~Benyamina, H.~Tran, I.~Yu,
  I.~Abdulmumin, I.~Johnson, I.~Gonzalez-Dios, J.~de~la Rosa, J.~Chim,
  J.~Dodge, J.~Zhu, J.~Chang, J.~Frohberg, J.~Tobing, J.~Bhattacharjee,
  K.~Almubarak, K.~Chen, K.~Lo, L.~V. Werra, L.~Weber, L.~Phan, L.~B. allal,
  L.~Tanguy, M.~Dey, M.~R. Muñoz, M.~Masoud, M.~Grandury, M.~Šaško,
  M.~Huang, M.~Coavoux, M.~Singh, M.~T.-J. Jiang, M.~C. Vu, M.~A. Jauhar,
  M.~Ghaleb, N.~Subramani, N.~Kassner, N.~Khamis, O.~Nguyen, O.~Espejel,
  O.~de~Gibert, P.~Villegas, P.~Henderson, P.~Colombo, P.~Amuok, Q.~Lhoest,
  R.~Harliman, R.~Bommasani, R.~L. López, R.~Ribeiro, S.~Osei, S.~Pyysalo,
  S.~Nagel, S.~Bose, S.~H. Muhammad, S.~Sharma, S.~Longpre, S.~Nikpoor,
  S.~Silberberg, S.~Pai, S.~Zink, T.~T. Torrent, T.~Schick, T.~Thrush,
  V.~Danchev, V.~Nikoulina, V.~Laippala, V.~Lepercq, V.~Prabhu, Z.~Alyafeai,
  Z.~Talat, A.~Raja, B.~Heinzerling, C.~Si, D.~E. Taşar, E.~Salesky, S.~J.
  Mielke, W.~Y. Lee, A.~Sharma, A.~Santilli, A.~Chaffin, A.~Stiegler, D.~Datta,
  E.~Szczechla, G.~Chhablani, H.~Wang, H.~Pandey, H.~Strobelt, J.~A. Fries,
  J.~Rozen, L.~Gao, L.~Sutawika, M.~S. Bari, M.~S. Al-shaibani, M.~Manica,
  N.~Nayak, R.~Teehan, S.~Albanie, S.~Shen, S.~Ben-David, S.~H. Bach, T.~Kim,
  T.~Bers, T.~Fevry, T.~Neeraj, U.~Thakker, V.~Raunak, X.~Tang, Z.-X. Yong,
  Z.~Sun, S.~Brody, Y.~Uri, H.~Tojarieh, A.~Roberts, H.~W. Chung, J.~Tae,
  J.~Phang, O.~Press, C.~Li, D.~Narayanan, H.~Bourfoune, J.~Casper, J.~Rasley,
  M.~Ryabinin, M.~Mishra, M.~Zhang, M.~Shoeybi, M.~Peyrounette, N.~Patry,
  N.~Tazi, O.~Sanseviero, P.~von Platen, P.~Cornette, P.~F. Lavallée,
  R.~Lacroix, S.~Rajbhandari, S.~Gandhi, S.~Smith, S.~Requena, S.~Patil,
  T.~Dettmers, A.~Baruwa, A.~Singh, A.~Cheveleva, A.-L. Ligozat,
  A.~Subramonian, A.~Névéol, C.~Lovering, D.~Garrette, D.~Tunuguntla,
  E.~Reiter, E.~Taktasheva, E.~Voloshina, E.~Bogdanov, G.~I. Winata,
  H.~Schoelkopf, J.-C. Kalo, J.~Novikova, J.~Z. Forde, J.~Clive, J.~Kasai,
  K.~Kawamura, L.~Hazan, M.~Carpuat, M.~Clinciu, N.~Kim, N.~Cheng, O.~Serikov,
  O.~Antverg, O.~van~der Wal, R.~Zhang, R.~Zhang, S.~Gehrmann, S.~Mirkin,
  S.~Pais, T.~Shavrina, T.~Scialom, T.~Yun, T.~Limisiewicz, V.~Rieser,
  V.~Protasov, V.~Mikhailov, Y.~Pruksachatkun, Y.~Belinkov, Z.~Bamberger,
  Z.~Kasner, A.~Rueda, A.~Pestana, A.~Feizpour, A.~Khan, A.~Faranak, A.~Santos,
  A.~Hevia, A.~Unldreaj, A.~Aghagol, A.~Abdollahi, A.~Tammour, A.~HajiHosseini,
  B.~Behroozi, B.~Ajibade, B.~Saxena, C.~M. Ferrandis, D.~McDuff,
  D.~Contractor, D.~Lansky, D.~David, D.~Kiela, D.~A. Nguyen, E.~Tan,
  E.~Baylor, E.~Ozoani, F.~Mirza, F.~Ononiwu, H.~Rezanejad, H.~Jones,
  I.~Bhattacharya, I.~Solaiman, I.~Sedenko, I.~Nejadgholi, J.~Passmore,
  J.~Seltzer, J.~B. Sanz, L.~Dutra, M.~Samagaio, M.~Elbadri, M.~Mieskes,
  M.~Gerchick, M.~Akinlolu, M.~McKenna, M.~Qiu, M.~Ghauri, M.~Burynok,
  N.~Abrar, N.~Rajani, N.~Elkott, N.~Fahmy, O.~Samuel, R.~An, R.~Kromann,
  R.~Hao, S.~Alizadeh, S.~Shubber, S.~Wang, S.~Roy, S.~Viguier, T.~Le,
  T.~Oyebade, T.~Le, Y.~Yang, Z.~Nguyen, A.~R. Kashyap, A.~Palasciano,
  A.~Callahan, A.~Shukla, A.~Miranda-Escalada, A.~Singh, B.~Beilharz, B.~Wang,
  C.~Brito, C.~Zhou, C.~Jain, C.~Xu, C.~Fourrier, D.~L. Periñán, D.~Molano,
  D.~Yu, E.~Manjavacas, F.~Barth, F.~Fuhrimann, G.~Altay, G.~Bayrak, G.~Burns,
  H.~U. Vrabec, I.~Bello, I.~Dash, J.~Kang, J.~Giorgi, J.~Golde, J.~D. Posada,
  K.~R. Sivaraman, L.~Bulchandani, L.~Liu, L.~Shinzato, M.~H. de~Bykhovetz,
  M.~Takeuchi, M.~Pàmies, M.~A. Castillo, M.~Nezhurina, M.~Sänger,
  M.~Samwald, M.~Cullan, M.~Weinberg, M.~D. Wolf, M.~Mihaljcic, M.~Liu,
  M.~Freidank, M.~Kang, N.~Seelam, N.~Dahlberg, N.~M. Broad, N.~Muellner,
  P.~Fung, P.~Haller, R.~Chandrasekhar, R.~Eisenberg, R.~Martin, R.~Canalli,
  R.~Su, R.~Su, S.~Cahyawijaya, S.~Garda, S.~S. Deshmukh, S.~Mishra,
  S.~Kiblawi, S.~Ott, S.~Sang-aroonsiri, S.~Kumar, S.~Schweter, S.~Bharati,
  T.~Laud, T.~Gigant, T.~Kainuma, W.~Kusa, Y.~Labrak, Y.~S. Bajaj,
  Y.~Venkatraman, Y.~Xu, Y.~Xu, Y.~Xu, Z.~Tan, Z.~Xie, Z.~Ye, M.~Bras,
  Y.~Belkada, and T.~Wolf, ``{BLOOM: A 176B-Parameter Open-Access Multilingual
  Language Model},'' 2023. [Online]. Available:
  \url{https://arxiv.org/abs/2211.05100}
\BIBentrySTDinterwordspacing

\bibitem{TGIS}
\BIBentryALTinterwordspacing
``{huggingface/text-generation-inference: Large Language Model Text Generation
  Inference},'' 2024. [Online]. Available:
  \url{https://github.com/huggingface/text-generation-inference}
\BIBentrySTDinterwordspacing

\bibitem{vLLM}
\BIBentryALTinterwordspacing
``{vllm-project/vllm: A high-throughput and memory-efficient inference and
  serving engine for LLMs},'' 2024. [Online]. Available:
  \url{https://github.com/vllm-project/vllm}
\BIBentrySTDinterwordspacing

\bibitem{SGLang-Arxiv}
\BIBentryALTinterwordspacing
L.~Zheng, L.~Yin, Z.~Xie, C.~Sun, J.~Huang, C.~H. Yu, S.~Cao, C.~Kozyrakis,
  I.~Stoica, J.~E. Gonzalez, C.~Barrett, and Y.~Sheng, ``{SGLang: Efficient
  Execution of Structured Language Model Programs},'' 2024. [Online].
  Available: \url{https://arxiv.org/abs/2312.07104}
\BIBentrySTDinterwordspacing

\bibitem{DaliBlog}
\BIBentryALTinterwordspacing
J.~A. Guirao, R.~Banas, K.~Łęcki, J.~Lisiecki, A.~Wolant, M.~Zientkiewicz,
  K.~Tokarski, and M.~Szołucha, ``{Rapid Data Pre-Processing with NVIDIA
  DALI},'' 2021. [Online]. Available:
  \url{https://developer.nvidia.com/blog/rapid-data-pre-processing-with-nvidia-dali/}
\BIBentrySTDinterwordspacing

\bibitem{OracleBlog}
\BIBentryALTinterwordspacing
P.~Valdria, ``{Accelerate AI and ML workloads with OCI, NVIDIA Magnum IO
  GPUDirect Storage, and IBM Storage Scale},'' 2023. [Online]. Available:
  \url{https://blogs.oracle.com/cloud-infrastructure/post/accelerate-ai-ml-workloads-oci-nvidia-ibm}
\BIBentrySTDinterwordspacing

\bibitem{Python-GIL}
\BIBentryALTinterwordspacing
{Sam Gross}, ``{PEP 703 --- Making the Global Interpreter Lock Optional in
  CPython},'' 2023, accessed: 2025-05-20. [Online]. Available:
  \url{https://peps.python.org/pep-0703/}
\BIBentrySTDinterwordspacing

\bibitem{SPIN-ATC17}
S.~Bergman, T.~Brokhman, T.~Cohen, and M.~Silberstein, ``{SPIN: Seamless
  Operating System Integration of Peer-to-Peer DMA Between SSDs and GPUs},'' in
  \emph{{2017 USENIX Annual Technical Conference (USENIX ATC '17)}}, 2017, pp.
  167--179.

\bibitem{P2PDMA-APSys20}
R.~Nakamura, Y.~Kuga, and K.~Akashi, ``{How beneficial is peer-to-peer DMA?}''
  in \emph{Proceedings of the 11th ACM SIGOPS Asia-Pacific Workshop on Systems
  (APSys '20)}, 2020, p. 25–32.

\bibitem{ColumnCache-BigData18}
T.~Yoshimura, T.~Chiba, and H.~Horii, ``{Column Cache: Buffer Cache for
  Columnar Storage on HDFS},'' in \emph{{2018 IEEE International Conference on
  Big Data (Big Data '18)}}, 2018, pp. 282--291.

\bibitem{TeraHeap-ASPLOS23}
I.~G. Kolokasis, G.~Evdorou, S.~Akram, C.~Kozanitis, A.~Papagiannis, F.~S.
  Zakkak, P.~Pratikakis, and A.~Bilas, ``Teraheap: Reducing memory pressure in
  managed big data frameworks,'' in \emph{Proceedings of the 28th ACM
  International Conference on Architectural Support for Programming Languages
  and Operating Systems, (ASPLOS '23)}, 2023, p. 694–709.

\bibitem{InMemArrShare-BigData23}
M.~Nozawa, S.~Imamura, and K.~Kono, ``Accelerating multilingual applications
  with in-memory array sharing,'' in \emph{{2023 IEEE International Conference
  on Big Data (BigData '23)}}, 2023, pp. 255--262.

\bibitem{FP8-Arxiv}
\BIBentryALTinterwordspacing
P.~Micikevicius, D.~Stosic, N.~Burgess, M.~Cornea, P.~Dubey, R.~Grisenthwaite,
  S.~Ha, A.~Heinecke, P.~Judd, J.~Kamalu, N.~Mellempudi, S.~Oberman,
  M.~Shoeybi, M.~Siu, and H.~Wu, ``{FP8 Formats for Deep Learning},'' 2022.
  [Online]. Available: \url{https://arxiv.org/abs/2209.05433}
\BIBentrySTDinterwordspacing

\bibitem{Quantization-Arxiv}
\BIBentryALTinterwordspacing
H.~Shen, N.~Mellempudi, X.~He, Q.~Gao, C.~Wang, and M.~Wang, ``Efficient
  post-training quantization with fp8 formats,'' 2024. [Online]. Available:
  \url{https://arxiv.org/abs/2309.14592}
\BIBentrySTDinterwordspacing

\bibitem{FP8Quantization-Arxiv}
\BIBentryALTinterwordspacing
A.~Kuzmin, M.~V. Baalen, Y.~Ren, M.~Nagel, J.~Peters, and T.~Blankevoort,
  ``{FP8 Quantization: The Power of the Exponent},'' 2024. [Online]. Available:
  \url{https://arxiv.org/abs/2208.09225}
\BIBentrySTDinterwordspacing

\end{thebibliography}
\bibliographystyle{IEEEtran}

\end{document}